\documentclass[floatfix,10pt,superscriptaddress,twocolumn,showpacs,showkeys,aps,prb,final]{revtex4-1}

\usepackage{graphicx}
\usepackage{dcolumn}
\usepackage{subfigure}
\usepackage{bm}
\usepackage{ifpdf,braket}
\usepackage{epstopdf}
\usepackage{amsmath}

\begin{document}

\title{Lattice Boltzmann method for bosons and fermions and the fourth order Hermite polynomial expansion}
\author{Rodrigo C. V. Coelho}
\affiliation{Universidade Federal do Rio de Janeiro, 68.528, Rio de
Janeiro - RJ, 21941-972, Brazil}
\author{Anderson Ilha}
\affiliation{Instituto Nacional de Metrologia, Normaliza\c{c}\~ao e
Qualidade Industrial, Duque de Caxias 25.250-020 RJ Brazil}
\author{M. M. Doria}
\affiliation{Universidade Federal do Rio de Janeiro, 68.528, Rio de
Janeiro - RJ, 21941-972, Brazil} \email{mmd@if.ufrj.br}
\author{R. M. Pereira}
\affiliation{Instituto Nacional de Metrologia,
Normaliza\c{c}\~ao e Qualidade Industrial, Duque de Caxias
25.250-020 RJ Brazil}
\author{Valter Yoshihiko Aibe}
\affiliation{Instituto Nacional de Metrologia, Normaliza\c{c}\~ao
e Qualidade Industrial, Duque de Caxias 25.250-020 RJ Brazil}
\date{\today}

\begin{abstract}
The Boltzmann equation with the Bhatnagar-Gross-Krook collision
operator is considered for the Bose-Einstein and Fermi-Dirac
equilibrium distribution functions. We show that the expansion of
the microscopic velocity in terms of Hermite polynomials must be
carried until the fourth order to correctly describe the energy
equation. The viscosity and thermal coefficients, previously
obtained by J.Y. Yang et al~\cite{Shi20089389,PhysRevE.79.056708}
through the Uehling-Uhlenbeck approach, are also derived here.
Thus the construction of a lattice Boltzmann method for the quantum fluid is possible provided that the Bose-Einstein and Fermi-Dirac equilibrium distribution functions are expanded until fourth order in the Hermite
polynomials.
\end{abstract}


%
%
%

\maketitle

\section{introduction}
One of the greatest achievements of the Boltzmann
equation~\cite{gilberto} is to determine the macroscopic
hydrodynamical equations of a fluid from a  phase space distribution
function, $f(\boldsymbol{\chi},\boldsymbol{x},t)$, which describes
the probability to find particles with microscopic velocity
$\boldsymbol{\chi}$ in position $\boldsymbol{x}$  at time $t$.
Nearly eighty years have passed since E. A. Uehling and G. E.
Uhlenbeck~\cite{PhysRev.43.552} solved the Boltzmann equation for
the quantum fluid approximately by determining the small correction
to the distribution function of non-interacting particles in case of
a weak interaction. From this solution they derived the macroscopic
hydrodynamical equations through the so-called Chapman-Enskog
analysis and obtained the viscosity, $\eta$, and the thermal
conductivity, $\kappa$, coefficients of the quantum fluid. The
Uehling-Uhlenbeck approach was later revisited by T. Nikuni and A.
Griffin~\cite{nikuni98} who derived the macroscopic hydrodynamic
equations, and the corresponding $\eta$ and $\kappa$ coefficients,
of a trapped Bose gas above the Bose-Einstein condensation with
damping. C. H. Lepienski and G. M. Kremer~\cite{lepienski96} also
determined these coefficients in case of  specific two-body
potentials, namely, Lennard-Jones and hard spheres. Recently the
Boltzmann equation was found applicable to describe the collective
oscillations of the two-dimensional Fermi
gas~\cite{PhysRevLett.108.070404}. All the above studies of the
quantum fluid take the assumption of a two-body collision operator,
as in the original Bolztmann equation. A simplifying assumption for
the collision operator was introduced in the fifties by
Bhatnagar-Gross-Krook (BGK)~\cite{bgk54}, who considered it just as
a simple drive to an {\it equilibrium distribution function} under a
single relaxation time $\tau$. Nevertheless only recently the
Uehling-Uhlenbeck approach was applied to solve the BGK-Boltzman
equation for the quantum fluid~\cite{Shi20089389,PhysRevE.79.056708}
by J.Y. Yang et al., who derived its $\eta$ and $\kappa$ coefficients.

In the late eighties a numerical scheme was formulated to solve the
Boltzmann equation with the BGK collision
term~\cite{succi,sukop-thorne,wolf-gladrow,mohamad} under the
assumption of a discrete phase space where both the microscopic
velocity  and the position are restricted to a discrete set of
values defined by a lattice. This method became widely known as the
lattice Boltzmann method (LBM) and is used to simulate fluids with
numerous advantages, such as easy implementation, inherent
parallelization, and flexible treatment of the boundary conditions.
The position space falls on a regular lattice where each point has a
discrete set of microscopic velocity vectors  that points towards a
selected set of nearest nodes. To this discrete set of directions we
associate the index $\alpha$, such that the microscopic velocities
become $\boldsymbol{\chi}_{\alpha}$. The neighbor points are reached
after a time $\Delta t$. The discreteness and rigidity of the
microscopic velocity in the LBM makes the distribution function also
become a discrete set, and instead of
$f(\boldsymbol{\chi},\boldsymbol{x},t)$, one has
$f_{\alpha}(\boldsymbol{x},t)$, fact that is of great numerical
advantage. Then the lattice BGK-Boltzmann equation is derived from
the continuous Boltzmann equation under a discretization
procedure~\cite{PhysRevE.56.6811},
\begin{eqnarray}\label{disboltzeq0}
&&f_\alpha(\boldsymbol{x}+ \boldsymbol{\chi}_\alpha\Delta t ,
t+\Delta
t)- f_\alpha (\boldsymbol{x}, t) = \nonumber \\
&&=- \frac{\Delta t}{\tau} \left[ f_\alpha(\boldsymbol{x}, t) -
f_\alpha^{(0)}(\boldsymbol{x}, t) \right],
\end{eqnarray}
which constantly drives the non-equilibrium distribution
$f_\alpha(\boldsymbol{x}, t)$ to the equilibrium distribution
function $f_\alpha^{(0)}(\boldsymbol{x}, t)$. Despite the tremendous
success of the LBM method to describe the mass and momentum
(Navier-Stokes) equations, the inclusion of a energy equation
remained a challenge for sometime. The energy equation is needed to
describe, for instance, the conversion of friction due to motion into heat, that increases the fluid temperature. This means that the transport of
matter by particle diffusion is intertwined with the advected
transport of enthalpy in such a way that the total energy is
conserved for a closed system. The macroscopic hydrodynamical
equations of the classical thermal compressible fluid are
well-known, and can be derived from general macroscopic principles,
as shown in the book of Landau \& Lifshitz~\cite{landau}, for
instance.

Many years after the development of the Uehling-Uhlenbeck approach,
H. Grad~\cite{grad49,PhysRevLett.78.243} devised another method to
solve the continuous Boltzmann equation based on a sequence of
approximations, obtained through the expansion of the distribution
function in terms of the microscopic velocity space, expressed as a
gaussian times a linear expansion in Hermite polynomials. Grad's
approach turned to be of paramount importance for the understanding
of the properties of the LBM. For this reason the Hermite polynomial
expansion method found a renewal of interest, such as in
Refs.~\onlinecite{PhysRevLett.80.65,FLM:409537}. Despite the
understanding brought by these references, the ingredients to
describe the thermal compressible classical fluid were still
missing, since the Hermite polynomial expansion was only carried
there until third order ($N=3$), which is just insufficient. It was
not until recently that the LBM for the thermal compressible fluid
with a single BGK relaxation time, as described in
Eq.(\ref{disboltzeq0}), was derived. Philippi et al.
(Ref.~\onlinecite{PhysRevE.73.056702}), Siebert et al.
(Ref.~\onlinecite{philippi07}), and Shan and Chen
(Ref.~\onlinecite{shan-chen07}), succeeded to show that the thermal
compressible properties of the classical fluid are correctly
described if the Hermite polynomial expansion is carried until
fourth order ($N=4$). Then the Chapman-Enskog
analysis~\cite{PhysRevE.73.056702,philippi07,shan-chen07}, applied
to the BGK-Boltzmann equation with the Maxwell-Boltzmann equilibrium
distribution function, gives the macroscopic hydrodynamical
equations for the mass, momentum and energy balance, as obtained in
the book of Landau \& Lifshitz~\cite{landau}.


In this paper we apply the $N=4$ order Hermite polynomial expansion
procedure to the quantum fluid and obtain its  macroscopic
hydrodynamical equations from a Chapman-Enskog analysis of the
BGK-Boltzmann equation. We conclude that the construction of an LBM
for the quantum fluid, similarly to the classical fluid, requires
the Hermite polynomial expansion be carried until $N=4$ order to reach
the correct energy equation, which besides the mass and momentum
equations, is also needed for a full description of friction-heating
processes. The coefficients $\eta$ and $\kappa$ of the BGK-Boltzmann
quantum fluid, obtained through  the Uehling-Uhlenbeck
approach~\cite{Shi20089389,PhysRevE.79.056708}, can only be
retrieved in the $N=4$ order, and not in lowest order, as shown
here. Therefore the halt of the Hermite polynomial expansion in
$N=3$ order, as done in Ref.~\onlinecite{PhysRevE.79.056708},
impairs the derivation of the correct energy equation of the quantum
fluid. Similarly the macroscopic hydrodynamical equations of
Ref.~\onlinecite{yang10} do not correctly describe the energy
balance equation because they are limited to $N=3$ order.
Ref.~\onlinecite{PhysRevE.79.056708} provides the general form of
the macroscopic hydrodynamical equations for the BGK-Boltzmann
quantum fluid, but not the formulation of these equations  in terms
of the local macroscopic fields, namely, the chemical potential
$\mu(\boldsymbol{x})$ (or the fugacity $z(\boldsymbol{x})$), the
local temperature $\theta(\boldsymbol{x})$, and the local velocity
$\boldsymbol{u}(\boldsymbol{x})$. In this paper we obtain the
macroscopic hydrodynamical equation in terms of the above
macroscopic variables.

We summarize the main achievements of this paper as follows. We
obtain the equilibrium distribution function of the Bose-Einstein
(BE) and Fermi-Dirac (FD) expanded until $N=4$ order in Hermite
polynomials, namely, Eq.(\ref{edf-cont}), or equivalently, Eq.(\ref{edf-dis}). We also obtain the Maxwell-Boltzmann (MB)
equilibrium distribution function by Taylor expansion, namely,
Eq.(\ref{fTaylor}). In the classical limit the $N=4$ BE-FD
equilibrium distribution function becomes  the Taylor expanded MB
equilibrium distribution function. All our results are obtained  in
$D$ dimensions, a feature that generalizes previous results
discussed in the literature~\cite{Shi20089389,PhysRevE.79.056708}.
Next we derive the macroscopic hydrodynamical equations for the
BE-FD cases, Eq.(\ref{navierstokeseq}) and Eq.(\ref{energy-3}), and
show that they become their MB counterparts,
Eq.(\ref{momentum-landau}) and Eq.(\ref{energy-landau}) in the
classical limit, where BE, FD and MB statistics coincide. The
viscosity and the thermal coefficients, given by
Eq.(\ref{eta_definition}) and Eq.(\ref{thermal_coeff}),
respectively, are the ones previously derived in
Ref.~\onlinecite{Shi20089389} by another method.
We show in Section~\ref{n=3} that the
$N=3$ Hermite expansion  yields an incorrect energy equation and
therefore this order is not suitable to describe either the
classical or the quantum fluid. The $N=4$ order quantum fluid
equations cannot be generally mapped into the classical ones for
arbitrary fugacity, and only at the classical limit. However this
mapping, as found in this paper, is always possible for the $N=3$.
This shows, once again, that the $N=3$ order can not give a correct
description of the quantum fluid. We find here an interesting
property of the macroscopic hydrodynamical equations of the quantum
fluid. To better understand this property we introduce the
notation of {\it pseudo} variables, represented by a {\it bar} on
the top of the corresponding variable. The property corresponds to the
mapping of the {\it pseudo} variables into their counterparts, which
leads the quantum into the classical equations. This mapping of the
quantum into the classical equations is not complete though, as a single term in the heat flow term of the energy equation spoils it, as seen in Eq.(\ref{tilq}).
Finally we explicitly construct a LBM for the quantum fluids using the corresponding $N=4$ order LBM scheme and obtain some numerical results in the limit of dilute quantum fluids. In this dilute limit the BE and FD are small
corrections to the MB statistics. Our numerical analysis is based on
some two-dimensional lattices~\cite{FLM:409537,PhysRevE.77.026707},
the so-called d2q17 and d2q37 lattices. Thus we are able to check
that indeed quantum fluid motion generates friction and the heat
created results into a temperature change such that the total energy
remains conserved during a time evolution process. Our numerical
study is restricted to an adiabatic process as there is no external contact or force. We treat a two-dimensional system
under periodic boundary conditions, thus without external borders
and with no external applied force, just an initial displacement
from equilibrium set by the initial conditions.

The paper is organized as follows. In section~\ref{equations} we
review the classical macroscopic hydrodynamical equations as
described by Landau \& Lifshitz's~\cite{landau} and also those
obtained from a Chapman-Enskog analysis of the $N=4$ order Hermite
polynomial
expansion~\cite{PhysRevE.73.056702,philippi07,shan-chen07}. We also
present in section~\ref{equations} the macroscopic hydrodynamical
equations of the quantum fluids and discuss several of their
aspects, such as the introduction of the pseudo variables and the
viscosity and thermal coefficients. The dimensionless units are also
introduced in section~\ref{equations}. In the next
section~\ref{hermite} the Hermite polynomial decomposition is
performed on the BE and FD equilibrium distribution functions. The
derivation of the MB equilibrium distribution function by Taylor
expansion is left for the appendix~\ref{appendix-Taylor}. Some
properties of the Hermite polynomials are discussed  in
appendix~\ref{appendix-hermite}. The passage from the continuum to
the discrete through the Gauss-Hermite quadrature is the subject of
section~\ref{quadrature}. The Chapman-Enskog analysis is performed
in  section~\ref{chapman-enskog} and the relations required to
determine several tensors are listed in the
appendix~\ref{usefulrelations}. To obtain the macroscopic
hydrodynamical equations one needs some position and time cross
derivative terms derived in
appendix~\ref{appendix-cross-derivative}. In section~\ref{n=3} we
obtain a no go theorem for the Hermite polynomial expansion to order
$N=3$. The dilute quantum fluid is treated in section~\ref{dilute}
and numerical results in this limit are studied in
section~\ref{numerical}.

There has been in the past years  attempts to construct a LBM for
the classical thermal compressible fluid starting from {\it ad hoc}
assumptions of the equilibrium distribution
function~\cite{PhysRevE.47.R2249,qian93,PhysRevE.50.2776,wolf-gladrow,PhysRevE.76.016702}.
This approach was never attempted for the quantum fluid.

\section{Macroscopic equations for classical and quantum fluids}\label{equations}
In this section we present the macroscopic hydrodynamical equations
of the quantum fluid obtained in this paper through the $N=4$
Hermite polynomial expansion of the BE and FD equilibrium
distribution functions. We express them almost entirely in terms of the
{\it pseudo} variables, with the exception of a single term in the heat flow.
From these equations we obtain the thermal coefficients, and
the classical MB limit. The actual derivation of these equations is
done in sections~\ref{hermite} and~\ref{chapman-enskog}, with the
help of appendices~\ref{usefulrelations}
and~\ref{appendix-cross-derivative}. In order to best convey our
results we review the derivation of  the macroscopic hydrodynamical
equations for the classical fluid, both from the point of view of
the macroscopic principles and of the N=4 Hermite polynomial expansion
of the Maxwell-Boltzmann equilibrium distribution
function~\cite{PhysRevE.73.056702,philippi07,shan-chen07}. Thus the
first two subsections are reviews while the last one contains our
original results.

\subsection{Derivation of the macroscopic equation from general principles}
For the ideal fluid (no viscosity) the equations describing the
conservation of mass, momentum and energy follow straightforwardly
from general macroscopic principles of thermodynamics and newtonian
mechanics, as well described, for instance,  in the book of Landau
\& Lifshitz~\cite{landau}. For the conservation of mass one obtains
that
\begin{eqnarray}\label{continuity}
\frac{\partial \rho}{\partial t} + \frac{\partial }{\partial x^i}
(\rho u^i)=0.
\end{eqnarray}
For the conservation of momentum one obtains Euler's equation,
\begin{eqnarray}\label{euler}
\frac{\partial \rho u^i}{\partial t} + \frac{\partial }{\partial
x^j}(p\delta^{ij}+\rho u^i u^j)=0,
\end{eqnarray}
and the law of conservation of energy is
\begin{eqnarray}\label{energia-ideal}
\frac{\partial \left (\rho\boldsymbol{u}^2/2+\rho\varepsilon
\right)}{\partial t} + \frac{\partial }{\partial
x^j}[u^j(\rho\boldsymbol{u}^2/2 +\rho\varpi)]=0,
\end{eqnarray}
where $p$, $\varepsilon$ and $\varpi$ describe the pressure, the
internal energy and the enthalpy of the fluid, $\varpi \equiv
\varepsilon+p/\rho$. These equations describe the time evolution of
the density, $\rho$, the macroscopic velocity, $u^i$, and the local
temperature, $\theta$, in some reduced units, which are explained in
this paper. The presence of viscosity and thermal conduction in a
fluid changes the equations for momentum and energy, which become
\begin{eqnarray}
&& \frac{\partial(\rho u^i)}{\partial t} + \frac{\partial}{\partial
x^j}\left( p\delta^{ij}+\rho u^i u^j-\sigma^{ij}\right) =0,
\label{momentum-landau}
\end{eqnarray}
and
\begin{eqnarray}
&&\frac{\partial\left(\rho\boldsymbol{u}^2/2+\rho\varepsilon\right)}{\partial
t}+\frac{\partial}{\partial x^j}\left[u^j\left(\rho\boldsymbol{u}^2/2 +\rho\varpi\right)+ \right .\nonumber \\
&&\left. +Q^j - u^k \sigma^{jk}\right]=0\label{energy-landau},
\end{eqnarray}
respectively. The momentum flux tensor becomes $p\delta^{ij}+\rho
u^i u^j-\sigma^{ij}$, thus is changed by the presence of the
viscosity stress tensor,
\begin{eqnarray}\label{eta-navierstokes}
\sigma^{ij}=\eta \left ( \frac{\partial u^i}{\partial x^j}+
\frac{\partial u^j}{\partial x^i} -\frac{2}{D}\delta^{ij}
\frac{\partial u^l}{\partial x^l} \right
)+\zeta\delta^{ij}\frac{\partial u^l}{\partial x^l}
\end{eqnarray}
where $\eta$ and $\zeta$ being the dynamic and volumetric
viscosities, respectively, and $D$ the number of space dimensions.
Similarly, the total energy flux becomes $u^j(\rho\boldsymbol{u}^2/2
+\rho\varpi)+  Q^j-u^k \sigma^{jk}$, where
\begin{eqnarray}\label{heat-flux0}
Q^j= -\kappa \frac{\partial\theta}{\partial x^j}
\end{eqnarray}
is the heat flux flow due to thermal conduction, and $\kappa$ is the
thermal conductivity. The presence of viscosity introduces
irreversible transfer of momentum within the fluid and
Eq.(\ref{momentum-landau}) is essentially the Navier-Stokes
equation. From the other side notice that Eq.(\ref{energy-landau})
shows that in a viscous fluid the friction that stems from the loss
of momentum results into an increase of temperature thus conserving
the total energy in the process.

For the sake of completeness we express the equations for momentum
and energy in a different fashion. obtained by simple manipulation
of Eqs. (\ref{continuity}),  (\ref{momentum-landau}) and
(\ref{energy-landau}). The resulting equations are expressed in
terms of the stress tensor
\begin{eqnarray}\label{stress-tensior0}
P^{i\,j} \equiv \rho \theta \delta^{i\,j}-\sigma^{i\,j}.
\end{eqnarray}
The momentum conservation equation becomes,
\begin{eqnarray}\label{momento5}
\rho\left( \frac{\partial}{\partial t}+u^j \frac{\partial}{\partial
x^j}\right)u^i+\frac{\partial}{\partial x^j}P^{i\,j}=0.
\end{eqnarray}
The energy conservation equation becomes,
\begin{eqnarray}\label{energia5}
&&\frac{\partial}{\partial t}\left(\frac{D}{2}\rho\theta\right)+
\frac{\partial}{\partial x^j}\left(\frac{D}{2}\rho\theta u^j \right)+\nonumber \\
&&+\frac{1}{2}\left( \frac{\partial u^i}{\partial x^j}+\frac{\partial u^j}{\partial x^i}\right)P^{i\,j}+\frac{\partial}{\partial x^i}Q^{i}= 0.
\end{eqnarray}

\subsection{Macroscopic equations from the Maxwell-Boltzmann distribution function}
The Hermite polynomial expansion of the Maxwell-Boltzmann
equilibrium distribution function carried until $N=4$
order~\cite{PhysRevE.73.056702,philippi07,shan-chen07}, and the
subsequent Chapman-Enskog analysis applied to the BGK-Boltzmann
equation, exactly reproduce the above equations obtained by general
macroscopic principles. Thus the equations for the density,
Eq.(\ref{continuity}), the macroscopic velocity,
Eq.(\ref{momentum-landau}), and the temperature,
Eq.(\ref{energy-landau}) also hold here, and so do
Eq(\ref{momento5}) and Eq.(\ref{energia5}). However thermodynamic
functions and coefficients acquire special features. The viscosity
stress tensor has a null volumetric viscosity ($\zeta=0$), and  the
dynamic viscosity is given by,
\begin{eqnarray}\label{eta-true}
\eta\equiv \rho \theta\tau\left(1 - \frac{\Delta t}{2\tau}\right),
\end{eqnarray}
and the thermal conductivity by,
\begin{eqnarray} \label{kappa-true}
\kappa \equiv \left(\frac{D+2}{2}\right)\rho \theta\tau\left(1 - \frac{\Delta t}{2\tau}\right).
\end{eqnarray}
such that $\kappa/\eta=(D+2)/2$.
Notice that these coefficients are local because of their $\theta$
and $\rho$ dependence. The pressure is
$p=\rho\,\theta$, the internal energy is, $\varepsilon \equiv
D\theta/2$, such that the enthalpy becomes , $\varpi \equiv (D+2)\theta/2$.

\subsection{Macroscopic equations from the Bose-Einstein and Fermi-Dirac distribution functions}
The $N=4$ Hermite polynomial expansion of the BE and FD equilibrium
distribution functions, followed by the Chapman-Enskog analysis done
in the BGK-Boltzmann equation, leads to the macroscopic
hydrodynamical equations for the quantum fluid in terms of the
temperature $\theta$, the velocity $u^i$, and the chemical potential
$\mu$. The fugacity can be used instead of the chemical potential,
and is defined as,
\begin{equation}\label{fugacity0}
z \equiv e^{\frac{\mu}{\theta}}.
\end{equation}
We find that the quantum equations are very similar to the classical
ones, provided that some auxiliary variables are introduced,
starting with the pseudo-temperature defined below.
\begin{eqnarray}\label{bartheta}
\bar \theta =  \theta \frac{g_{\frac{D}{2}+1}(z)}{g_{\frac{D}{2}}(z)}
\end{eqnarray}
where,
\begin{eqnarray}\label{function-g}
g_{\nu}(z)\equiv \frac{1}{\Gamma(\nu)}\int^{\infty}_{0}\frac{x^{\nu-1}\,dx}{z^{-1}e^x\pm 1}.
\end{eqnarray}
The $+$ and $-$ signs in the denominator correspond to FD and BE
statistics, respectively. Indeed the MB limit is retrieved from the
FD and MB statistics in the limit that  $z^{-1}e^x\pm 1 \approx
z^{-1}e^x$, thus the term $\pm 1$ becomes irrelevant in
the denominator. Since $\Gamma(\nu)=\int^{\infty}_{0}e^{-x}
x^{\nu-1}\,dx$, one obtains that $g_{\nu}(z)=z$ for the MB limit.
Thus we conclude that in the classical limit the pseudo-temperature
is the temperature itself. The BE-FD density is not an independent
parameter, as in the MB case, but a function of the the temperature
$\theta$ and the fugacity $z$,
\begin{eqnarray}\label{rhodef}
\rho=(2\pi\theta)^{\frac{D}{2}} g_{\frac{D}{2}}(z).
\end{eqnarray}
Based on the above definitions of $\rho$ and $\bar \theta$ we
introduce the pseudo viscosity stress tensor
\begin{equation}\label{sigma_definition}
{\bar \sigma}^{ij} \equiv \bar \eta \left( \frac{\partial u^i}{\partial x^j} + \frac{\partial u^j}{\partial x^i} - \frac{2}{D} \delta^{ij} \right),
\end{equation}
which depends on the pseudo dynamic viscosity defined as,
\begin{equation}\label{eta_definition}
\bar \eta \equiv \rho\bar \theta \tau \left( 1-\frac{\Delta t}{2\tau} \right).
\end{equation}
We also define the pseudo thermal conductivity,
\begin{eqnarray}\label{kappa_definition}
\bar \kappa \equiv \frac{D+2}{2} \rho \bar \theta \tau \left( 1-\frac{\Delta
t}{2 \tau} \right).
\end{eqnarray}
Their ratio is constant, $\bar \kappa/\bar \eta = (D+2)/2$, meaning
that, like in the MB case,  thermal and mechanical transfers are
done by the same distribution and at the same rate, as there is only
one BGK time relaxation parameter.

The conservation of mass is given by Eq.(\ref{continuity}) and the
momentum balance equation for quantum fluids is described by,
\begin{equation} \label{navierstokeseq}
\frac{\partial}{\partial t}(\rho u^i) + \frac{\partial}{\partial
x^j} \left[ \rho \left( \bar \theta   \delta^{ij} + u^i u^j\right)
\right] - \frac{\partial {\bar \sigma}^{ij}}{\partial x^j} =0
\end{equation}
Comparison between Eq.(\ref{momentum-landau}) and
Eq.(\ref{navierstokeseq}) shows that the classical and the quantum
equations are the same provided that the true temperature $\theta$
is mapped into the pseudo temperature, $\bar \theta$. The same does
not hold for the energy equations, namely, Eq.(\ref{energy-landau})
and Eq.(\ref{energy-3}), as the dependence of the quantum  case in
the true temperature cannot collapse into the pseudo temperature.
The fugacity $z$ explicitly appears there through the function
$g(z)$, as seen below.
\begin{eqnarray}\label{energy-3}
&&\frac{\partial}{\partial t} \left( \frac{\rho}{2} \boldsymbol{u}^2 + \frac{\rho}{2}D \bar \theta   \right)+ \frac{\partial}{\partial x^j} \left[\left(\frac{\rho}{2}\boldsymbol{u}^2 + \frac{\rho}{2}\bar \theta (D+2) \right)u^j + \right . \nonumber \\
&&\left .+{\tilde Q}^j - u^k {\bar \sigma}^{j\,k}\right]=0,
\end{eqnarray}
where the true heat flux vector is given by,
\begin{eqnarray}\label{tilq}
{\tilde Q}^j = {\bar Q}^j - \frac{\partial}{\partial x^j}\left [ \bar \kappa \left( g(z)-1 \right)\bar \theta\right], \, {\bar Q}^{j} \equiv -\bar \kappa \frac{\partial}{\partial x^i}\bar \theta.
\end{eqnarray}
Thus the writing of the energy equation solely in terms of pseudo
variables is spoiled just by the extra term, that must be added to
the pseudo heat flux vector, ${\bar Q}^{j}$, to obtain the true one,
${\tilde Q}^j$. The function $g(z)$ is defined as,
\begin{eqnarray}\label{g(z)}
g(z) \equiv \frac{g_{\frac{D}{2}}(z)\cdot g_{\frac{D}{2}+2}(z)}{\left(g_{\frac{D}{2}+1}(z)\right)^2}.
\end{eqnarray}
In conclusion we find remarkable that the replacement of  ${\tilde
Q}^j$ by ${\bar Q}^j$  in Eq.(\ref{energy-3}) renders the mass,
momentum and energy equations for the quantum and classical fluids
formally identical, provided that the density and the temperature
are replaced by Eqs.(\ref{rhodef}) and (\ref{bartheta}),
respectively. Another way to see this connection is by noticing that
for the quantum problem all the explicit and {\it linear} dependence
in the temperature appears through the pseudo temperature. This is
the reason why the mass and momentum equations of quantum and
classical cases are formally equivalent, but not the energy
equation. The quantum energy equation has an extra {\it quadratic}
dependence on the temperature brought by the function $g(z)-1$, as
discussed in the coming sections. There is a link between this
quadratic behavior in the temperature and the $g(z)$ function, as
shown by the identity below.
\begin{eqnarray}\label{theta2}
 \theta^2\frac{g_{\frac{D}{2}+2}(z)}{g_{\frac{D}{2}}(z)}={\bar \theta}^2 g(z).
\end{eqnarray}
The quantum counterparts of the classical momentum and energy
equations, namely, Eqs.(\ref{momento5}) and (\ref{energia5}), are
given by
\begin{eqnarray}\label{momento-pq}
\rho\left( \frac{\partial}{\partial t}+u^j \frac{\partial}{\partial x^j}\right)u^i+\frac{\partial}{\partial x^j}{\bar P}^{i\,j}=0,
\end{eqnarray}
and,
\begin{eqnarray}\label{energia-pq}
&&\frac{\partial}{\partial t}\left(\frac{D}{2}\rho\bar\theta\right)+
\frac{\partial}{\partial x^j}\left(\frac{D}{2}\rho\bar\theta u^j \right)+\nonumber \\
&&+\frac{1}{2}\left( \frac{\partial u^i}{\partial x^j}+\frac{\partial u^j}{\partial x^i}\right){\bar P}^{i\,j}+\frac{\partial}{\partial x^i}{\tilde Q}^{i}= 0,
\end{eqnarray}
respectively. We also define the pseudo stress tensor,
\begin{eqnarray}\label{pseudo-stress-tensior}
{\bar P}^{i\,j} \equiv \rho \bar \theta \delta^{i\,j}-\bar \sigma^{i\,j}.
\end{eqnarray}

We obtain the true thermal conductivity coefficient,
$\kappa_{\theta}$, and the chemical potential coefficient,
$\kappa_{\mu}$, through their definition:
\begin{eqnarray}
{\tilde Q}^j = -\kappa_{\theta} \frac {\partial \theta}{\partial x^j} - \kappa_{\mu} \frac {\partial\mu}{\partial x^j}.
\end{eqnarray}
To derive the above coefficients, we write the pseudo heat flux
vector as,
\begin{eqnarray}\label{Qaux}
{\tilde Q}^j = \bar \theta \frac {\partial\bar \kappa}{\partial x^j} - \frac{\partial}{\partial x^j}\left [ \bar \kappa  g(z)\bar \theta\right].
\end{eqnarray}
Firstly derive the function $g_{\mu}(z)$ of Eq.(\ref{function-g}):
\begin{eqnarray}
\frac{\partial g_{\mu}(z)}{\partial x^j}=g_{\mu-1}(z)\left ( -\frac{\mu}{\theta}\frac{\partial \theta^2}{\partial x^j}+\frac{1}{\theta}\frac{\partial \mu}{\partial x^j}\right ),\end{eqnarray}
which leads to,
\begin{eqnarray}
\frac{\partial \bar \kappa}{\partial x^j}=\rho \tau' \left \{ \left [ \left( \frac{D}{2}+1 \right )\frac{g_{\frac{D}{2}+1}(z)}{g_{\frac{D}{2}}(z)}-\frac{\mu}{\theta}\right ]\frac{\partial \theta}{\partial x^j} +\frac{\partial \mu}{\partial x^j}\right \}, \nonumber \\
\end{eqnarray}
where $\tau'\equiv (D/2+1)\tau( 1-\Delta t/2\tau)$. One also obtains that,
\begin{eqnarray}
&&\frac{\partial \left( \bar \kappa \bar \theta g(z)\right)}{\partial x^j}=\rho \tau'\left\{\left[\left(\frac{D}{2}+2\right)\theta \frac{g_{\frac{D}{2}+2}(z)}{g_{\frac{D}{2}}(z)} -\mu \frac{g_{\frac{D}{2}+1}(z)}{g_{\frac{D}{2}}(z)}\right] \cdot \right . \nonumber \\
&&\left . \cdot \frac{\partial \theta}{\partial x^j}+ \theta \frac{g_{\frac{D}{2}+1}(z)}{g_{\frac{D}{2}}(z)} \frac{\partial \mu}{\partial x^j} \right\}.
\end{eqnarray}
The last two equations are  introduced in Eq.(\ref{Qaux}) to obtain
the coefficients $\kappa_{\theta}$ and  $\kappa_{\mu}$:
\begin{eqnarray}
&&\kappa_{\theta} = \kappa \left [ \left (\frac{D}{2}+2 \right) \frac{g_{\frac{D}{2}+2}(z)}{g_{\frac{D}{2}+1}(z)}- \left (\frac{D}{2}+1 \right) \frac{g_{\frac{D}{2}+1}(z)}{g_{\frac{D}{2}}(z)}\right ] \nonumber \\
\label{thermal_coeff}\\
&&\kappa_{\mu} = 0, \label{mu_coeff}
\end{eqnarray}
where $\kappa$ is the MB (classical) thermal conductivity, given by
Eq.(\ref{kappa-true}). These are the coefficients obtained in
Ref.~\onlinecite{Shi20089389} for the BGK-Boltzmann equation using
the Uehling-Uhlenbeck approach.

\section{Expansion of the quantum equilibrium distribution functions by Hermite polynomials to $N=4$ order} \label{hermite}
Here we expand the quantum equilibrium distribution functions until
$N=4$ order of Hermite polynomials. Consider the BE and FD
distribution functions, given by,
\begin{equation}
f^{(0)}(\boldsymbol{\chi})=\frac{1}{\exp\left[\frac{\left(\boldsymbol{\chi}-\boldsymbol{v}\right)^2m}{2k_BT}-\frac{\mu'}{k_BT}
\right]\pm1},
\end{equation}
which depend on the microscopic velocity $\boldsymbol{\chi}$ and
other three locally defined quantities, namely, the temperature
$T(\boldsymbol{x})$, the macroscopic velocity
$\boldsymbol{v}(\boldsymbol{x})$ and the chemical potential
$\mu'(\boldsymbol{x})$.

We define the following dimensionless quantities, based on a
reference temperature $T_r$, and a reference velocity:
\begin{eqnarray}\label{reduced-velocity}
c_r \equiv \sqrt{\frac{kT_r}{m}},
\end{eqnarray}
both not present in the original BE-FD equilibrium distribution functions.
Then the dimensionless temperature is
\begin{equation}
\theta\equiv\frac{T}{T_r},
\end{equation}
and the microscopic and macroscopic dimensionless velocities are defined as
\begin{eqnarray}
&&\boldsymbol{\xi} \equiv \frac{\boldsymbol{\chi}}{c_r}, \,\,\mbox{and}\\
&&\boldsymbol{u} \equiv \frac{\boldsymbol{v}}{c_r}.
\end{eqnarray}
The dimensionless chemical potential is,
\begin{equation}
\mu \equiv \frac{\mu'}{k_B T_r},
\end{equation}
and the fugacity can be expressed in two ways,
\begin{equation}
z\equiv e^{\frac{\mu'}{k_B T}} = e^{\frac{\mu}{\theta}}.
\end{equation}
The BE-FD function expressed in terms of dimensionless parameters becomes,
\begin{equation}\label{fd-dimensionless}
f^{(0)}(\boldsymbol{\xi})=\frac{1}{z^{-1}\exp\left[ \frac{\left(\boldsymbol{\xi}-\boldsymbol{u}\right)^2}{2\theta} \right]\pm 1}.
\end{equation}
The first three moments of the BE-FD function, obtained by
integrating over the microscopic velocity $\boldsymbol{\chi}$, give
that,
\begin{eqnarray} \label{moments}
&& n(\boldsymbol{x})=m\left(\frac{m}{2\pi \hbar}\right)^D\int d^D \boldsymbol{\chi}\; f^{0}(\boldsymbol{\chi}), \\
&& \boldsymbol{v}(\boldsymbol{x}) =\frac{1}{\rho(\boldsymbol{x})}\int
d^D \boldsymbol{\chi} \; \boldsymbol{\chi} f^{0}(\boldsymbol{\chi}), \, \mbox{and} \\
&& \frac{D}{2} k_B \bar T(\boldsymbol{x}) =\frac{1}{\rho(\boldsymbol{x})}\int d^D \boldsymbol{\chi}\;\frac{1}{2} m\left[\boldsymbol{\chi}-\boldsymbol{v}(\boldsymbol{x})\right]^2
f^{0}(\boldsymbol{\chi}).
\end{eqnarray}
We define the pseudo temperature $\bar T$ because the third moment
is the internal energy, $\varepsilon= D k_B\bar T/2$, and in the
classical case we know that $\bar T = T$. Notice that the density
$n$ has the dimension of a $D$ dimensional space density, as
expected, namely, of $mass/(length)^D$,  because $m v/2\pi \hbar$
has the dimension of $length^{-1}$. Thus through the following
constant, which has the dimension of density,
\begin{eqnarray}
n_0\equiv m\left(\frac{m c_r}{2\pi \hbar}\right)^D.
\end{eqnarray}
By taking averages over the quantum equilibrium distribution
function we obtain the dimensionless density $\rho$ of
Eq.(\ref{rhodef}), the macroscopic velocity $\boldsymbol{u}$, and
the pseudo-temperature $\bar \theta$ of Eq.(\ref{bartheta}):
\begin{eqnarray} \label{moments1}
&&\rho(\boldsymbol{x})\equiv \frac{n(\boldsymbol{x})}{n_0}=\int d^D \boldsymbol{\xi}\; f^{(0)}(\boldsymbol{\xi}), \\
&&\boldsymbol{u}(\boldsymbol{x}) =\frac{1}{\rho(\boldsymbol{x})}\int
d^D \boldsymbol{\xi} \; \boldsymbol{\xi} f^{(0)}(\boldsymbol{\xi}), \, \mbox{and} \\
&&\frac{D}{2} \bar \theta(\boldsymbol{x}) =\frac{1}{\rho(\boldsymbol{x})}\int d^D \boldsymbol{\xi}\;\frac{1}{2} \left[\boldsymbol{\xi}-\boldsymbol{u}(\boldsymbol{x})\right]^2
f^{(0)}(\boldsymbol{\xi}).
\end{eqnarray}
The explicit calculation of these moments are done below.

The N$^{\mbox{th}}$ order Hermite polynomial is defined by the
Rodrigues' formula,
\begin{equation}
H^{i_1i_2...i_N}(\boldsymbol{\xi}) = \frac{(-1)^N}{\omega(\boldsymbol{\xi})}
\frac{\partial^N \omega(\boldsymbol{\xi})}{\partial \xi^{i_1}
\xi^{i_2}... \xi^{i_N}}, \label{Hpoly}
\end{equation}
where,
\begin{equation}
\omega(\boldsymbol{\xi}) \equiv  \frac{1}{(2\pi)^{\frac{D}{2}}}
\exp{\left(- \frac{\boldsymbol{\xi}^2}{2}\right)}, \label{omega0}
\end{equation}
is the gaussian function.
The orthonormality  of the Hermite
polynomials, $H^{i_1i_2...i_N}(\boldsymbol{\xi})$, is given by,
\begin{eqnarray} \label{hermiteortho}
&&\int d^D \boldsymbol{\xi}\, \omega(\boldsymbol{\xi})
H^{i_1i_2...i_N}(\boldsymbol{\xi})
H^{j_1j_2...j_M}(\boldsymbol{\xi}) = \\
&&= \delta^{N M} ( \delta^{i_1j_1}\delta^{i_2j_2}...\delta^{i_Nj_N}
+ \mbox{all permutations of j's}). \nonumber
\end{eqnarray}
 We seek the decomposition of the BE-FD function $f^{(0)}$ of Eq.(\ref{fd-dimensionless}) in powers of Hermite polynomials.
\begin{equation}\label{fhermite}
f^{(0)}(\boldsymbol{x}, \boldsymbol{\xi})= \omega(\boldsymbol{\xi}) \sum^{\infty}_{N=0}
\frac{1}{N!}a^{i_1i_2...i_N}(\boldsymbol{x})
H^{i_1i_2...i_N}(\boldsymbol{\xi}).
\end{equation}
Notice that this decomposition splits the dependence on the
dimensionless microscopic velocity, which falls in the Hermite
polynomials, from the other variables, since the coefficients of
this expansion carry all the information about the local macroscopic
fields. For this reason the following notation is employed,
$a^{i_1\,i_2...i_N}(\boldsymbol{x})\equiv
a^{i_1\,i_2...i_N}[z(\boldsymbol{x}),\boldsymbol{u}(\boldsymbol{x}),\theta(\boldsymbol{x})]$.
Applying the orthonormality of the Hermite polynomials one obtains
an expression for the coefficients:
\begin{equation}\label{coeff}
a^{i_1i_2...i_N}(\boldsymbol{x}) = \int f^{(0)}(\boldsymbol{x}, \boldsymbol{\xi})
H^{i_1i_2...i_N}(\boldsymbol{\xi}) d^D\boldsymbol{\xi}
\end{equation}
More details about properties of the Hermite polynomials are given
in Appendix \ref{appendix-hermite}. Notice that the BE-FD function
is even in the difference between the macroscopic and the
microscopic velocities,
$f^{(0)}(\boldsymbol{x},\boldsymbol{\xi-u})=f^{(0)}(\boldsymbol{x},\boldsymbol{u-\xi})$,
a helpful property to compute the coefficients of Eq.(\ref{coeff}).
Then the coefficient associated to any odd Hermite polynomial
vanishes since $\int d^D
\boldsymbol{\xi}\,f(\boldsymbol{\xi-u})H^{i_1\ldots
i_{2L+1}}(\boldsymbol{\xi-u}) = 0$. A discussion of such properties
is done in the appendix~\ref{appendix-hermite}. We introduce the
variable $\boldsymbol{\eta}=\boldsymbol{\xi}-\boldsymbol{u}$ and
write the BE-FD distribution function as
$f(\boldsymbol{\eta})=1/\left(z^{-1}e^{\frac{\eta^2}{2\theta}}\pm
1\right)$ to express the coefficients as,
\begin{equation}
a^{i_1i_2...i_N}=\int f^{(0)}(\boldsymbol{\eta}) H^{i_1i_2...i_N}(\boldsymbol{\xi}) d^D\boldsymbol{\eta}.
\end{equation}
\\ \noindent {\bf Coefficient N=0} - The integration over a spherical shell in D dimensions is $d^D\boldsymbol{\eta} =\frac{D\pi^{\frac{D}{2}}}{\Gamma\left(\frac{D}{2}+ 1\right)}\eta^{D-1}d\eta$ and the lowest Hermite coefficient becomes,
\begin{eqnarray}
a = \int \frac{H d^D\boldsymbol{\eta}}{z^{-1}e^{\frac{\eta^2}{2\theta}}\pm 1} =\frac{D\pi^{\frac{D}{2}}}{\Gamma\left(\frac{D}{2}\pm 1\right)}\int^\infty_0 \frac{\eta^{D-1}d\eta}{z^{-1}e^{\frac{\eta^2}{2\theta}}\pm 1 }.
\end{eqnarray}
since $H=1$. Defining $x=\eta^2/2\theta$, one obtains that,
\begin{eqnarray}
a= (2\pi)^{\frac{D}{2}} \theta^{\frac{D}{2}} g_{\frac{D}{2}}(z)
\end{eqnarray}
where in the last line we have introduced the function $g_{\nu}(z)$ of Eq.(\ref{function-g}). This coefficient is the density itself, $a=\rho=\int d^D \boldsymbol{\xi}\; f^{(0)}(\boldsymbol{\xi})$, as given by Eq.(\ref{rhodef}).
\\
\\ \noindent {\bf Coefficient N=1} - To obtain this coefficient take Eq.(\ref{hermite-with-u-1}) such that,
\begin{eqnarray}
&&a^i=\int f^{(0)}(\boldsymbol{\eta}) H^i(\boldsymbol{\xi}) d^D\boldsymbol{\eta} = \nonumber \\ && =\int f^{(0)}(\boldsymbol{\eta}) \left[H^i(\boldsymbol{\eta}) + H u^i \right ] d^D\boldsymbol{\eta}  =  u^i \int f^{(0)}(\boldsymbol{\eta}) d^D\boldsymbol{\eta} = \nonumber \\
&&=  u^i \rho
\end{eqnarray}
as the first integral vanishes because it is odd. Using analogous
arguments, we find the next coefficients.
\\
\\ \noindent{\bf Coefficient N=2} - One gets from Eq.(\ref{hermite-with-u-2}) that,
\begin{eqnarray}
&&a^{ij} = \int f^{(0)}(\boldsymbol{\eta}) H^{ij}(\boldsymbol{\xi})d^D\boldsymbol{\eta} = \int f^{(0)}(\boldsymbol{\eta}) \left [ H^{ij}(\boldsymbol{\eta})+ \right . \nonumber\\
&&\left . + u^i  H^j(\boldsymbol{\eta}) + u^j H^i(\boldsymbol{\eta})+u^i u^j H \right ] d^D\boldsymbol{\eta}= \nonumber \\
&&=\int f^{(0)}(\boldsymbol{\eta}) H^{ij} (\boldsymbol{\eta}) d^D\boldsymbol{\eta} + \rho u^i u^j.
\end{eqnarray}
The terms proportional to $u^j$ vanish because of the odd integrals and it remains to calculate the integrals below.
\begin{eqnarray}
&&\int f^{(0)}(\boldsymbol{\eta}) \eta^i \eta^j d^D\boldsymbol{\eta} = \frac{\delta^{ij}}{D} \int f^{(0)}(\boldsymbol{\eta}) \eta^2 d^D\boldsymbol{\eta} = \nonumber \\ && = \theta \rho \frac{g_{\frac{D}{2}+1}(z)}{g_{\frac{D}{2}}(z)} \delta^{ij}. \label{int2eta}
\end{eqnarray}
The other integral to compute is,
\begin{equation}
\int f^{(0)}(\boldsymbol{\eta}) H^{ij}(\boldsymbol{\eta}) d^D\boldsymbol{\eta} = \rho\delta^{ij}\left( \theta \frac{g_{\frac{D}{2}+1}(z)}{g_{\frac{D}{2}}(z)} -1 \right).
\end{equation}
Introducing the pseudo temperature of Eq.(\ref{bartheta}), the second order coefficient becomes,
\begin{equation}
a^{ij} = \rho \left[ \delta^{ij}\left( \bar \theta -1 \right)+ u^i u^j \right].
\end{equation}
Hereafter we directly use Eq.(\ref{bartheta}) to shorten the notation.
\\ \noindent {\bf Coefficient N=3} - One gets from Eq.(\ref{hermite-with-u-3}) that,
\begin{eqnarray}
&& a^{ijk} = \int f^{(0)}(\boldsymbol{\eta}) H^{ijk}(\boldsymbol{\xi}) d^D\boldsymbol{\eta} = \nonumber \\
&&= \int f^{(0)}(\boldsymbol{\eta}) \left[H^{ijk}(\boldsymbol{\eta})+ u^i H^{jk}(\boldsymbol{\eta})+u^j H^{ik}(\boldsymbol{\eta}) + u^k H^{ij}(\boldsymbol{\eta})+ \right. \nonumber \\
&&\left. + u^i u^j H^k(\boldsymbol{\eta}) + u^i u^k H^j(\boldsymbol{\eta}) + u^j u^k H^i(\boldsymbol{\eta}) + u^i u^j u^k H \right ] d^D\boldsymbol{\eta} = \nonumber \\
&&= \rho \left[(u^i \delta^{jk}+ u^j \delta^{ik}+ u^k \delta^{ij}) \bar \theta  + u^i u^j u^k \right].\nonumber \\
\end{eqnarray}
Terms proportional to $u^iu^j$ vanish because they are proportional to
integrals over a Hermite polynomial of odd order.
\\
\\ \noindent {\bf Coefficient N=4} - One gets from Eq.(\ref{hermite-with-u-4}) that,
\begin{eqnarray}
&&a^{ijkl} = \int f^{(0)}(\boldsymbol{\eta}) H^{ijkl}(\boldsymbol{\xi}) d^D\boldsymbol{\eta} = \int f^{(0)}(\boldsymbol{\eta}) \left[H^{ijkl}(\boldsymbol{\eta}) +\right .\nonumber \\
&&+ u^i H^{jkl}(\boldsymbol{\eta}) + u^j H^{ijk}(\boldsymbol{\eta}) + u^k H^{ijl}(\boldsymbol{\eta}) + u^l H^{ijk}(\boldsymbol{\eta}) + \nonumber \\
&& +u^i u^j H^{kl}(\boldsymbol{\eta}) + u^i u^k H^{jl}(\boldsymbol{\eta})+ u^i u^l H^{jk}(\boldsymbol{\eta})+ u^k u^lH^{ij}(\boldsymbol{\eta}) + \nonumber \\
&&+ u^j u^l H^{ik}(\boldsymbol{\eta}) + u^j u^k H^{il}(\boldsymbol{\eta}) +u^i u^j u^k H^l(\boldsymbol{\eta}) + u^i u^j u^l H^k(\boldsymbol{\eta}) + \nonumber \\
&&\left. + u^j u^k u^l H^i(\boldsymbol{\eta})+ u^i u^k u^l H^j(\boldsymbol{\eta}) +  u^i u^j u^k u^l H \right] d^D\boldsymbol{\eta}.
\end{eqnarray}
Because of the odd integrals the terms proportional to $u^j$ and $u^i u^j u^k $ do not contribute, and so,
\begin{eqnarray}
&&a^{ijkl}= \int f^{(0)}(\boldsymbol{\eta}) H^{ijkl}(\boldsymbol{\eta})d^D\boldsymbol{\eta} + \nonumber \\
&&+ u^i u^j \int f^{(0)}(\boldsymbol{\eta}) H^{kl}(\boldsymbol{\eta})d^D\boldsymbol{\eta} + u^i u^k \int f^{(0)}(\boldsymbol{\eta})H^{jl}(\boldsymbol{\eta}) d^D\boldsymbol{\eta} + \nonumber \\
&&+u^i u^l\int f^{(0)}(\boldsymbol{\eta}) H^{jk}(\boldsymbol{\eta})d^D\boldsymbol{\eta} + u^k u^l \int f^{(0)}(\boldsymbol{\eta})H^{ij}(\boldsymbol{\eta}) d^D\boldsymbol{\eta} + \nonumber \\
&&+ u^j u^l \int f^{(0)}(\boldsymbol{\eta})H^{ik}(\boldsymbol{\eta}) d^D\boldsymbol{\eta} + u^j u^k \int f^{(0)}(\boldsymbol{\eta}) H^{il}(\boldsymbol{\eta}) d^D\boldsymbol{\eta} +\nonumber \\
&&+ \rho u^i u^j u^k u^l.
\end{eqnarray}
Defining the first integral as,
\begin{eqnarray}
&& I_1 \equiv \int f^{(0)}(\boldsymbol{\eta})H^{ijkl}(\boldsymbol{\eta}) d^D\boldsymbol{\eta} =\int \eta^i \eta^j \eta^k \eta^l f^{(0)}(\boldsymbol{\eta})d^D\boldsymbol{\eta}- \nonumber \\
&& -\delta^{kl} \int \eta^i \eta^j f^{(0)}(\boldsymbol{\eta}) d^D\boldsymbol{\eta} - \delta^{jl} \int \eta^i \eta^k f^{(0)}(\boldsymbol{\eta})d^D\boldsymbol{\eta} \nonumber - \\
&& - \delta^{jk} \int \eta^i \eta^l f^{(0)}(\boldsymbol{\eta}) d^D\boldsymbol{\eta}-\delta^{il} \int \eta^j \eta^k f^{(0)}(\boldsymbol{\eta}) d^D\boldsymbol{\eta}-\nonumber \\
&&-\delta^{ik} \int \eta^j \eta^l f^{(0)}(\boldsymbol{\eta}) d^D\boldsymbol{\eta} -\delta^{ij} \int \eta^k \eta^l f^{(0)}(\boldsymbol{\eta}) d^D\boldsymbol{\eta} + \nonumber \\
&&+ \rho \delta^{ijkl},
\end{eqnarray}
where the last symbol is defined in Eq.(\ref{delta-4}).
Then use Eq.(\ref{int2eta}) and call the first integral as $I_2$,
\begin{eqnarray}
I_1 = I_2 + \rho \left( 1-2 \bar\theta  \right) \delta^{ijkl}.
\end{eqnarray}
where,
\begin{equation}
I_2 \equiv \int \frac{\eta^i \eta^j \eta^k \eta^l}{z^{-1}e^{\frac{\eta^2}{2\theta}}\pm 1}d^D\boldsymbol{\eta} = \rho\theta^2  \frac{g_{\frac{D}{2}+2}(z)}{g_{\frac{D}{2}}(z)} \delta^{ijkl}.
\end{equation}
We stress that only the $N=4$ coefficient contains contributions
proportional to $\theta^2$. Thus an expansion of the BE-FD function
limited to the $N=3$ coefficient has the temperature only inside the
pseudo temperature. At this point we introduce the notation defined
in Eq.(\ref{theta2}) which brings the function $g(z)$ of
Eq.(\ref{g(z)}) into the $N=4$ coefficient,
\begin{equation}
I_1 = \rho \left( {\bar \theta}^2 g(z)  -2\bar \theta +1 \right) \delta^{ijkl}.
\end{equation}
Finally, we obtain that the fourth coefficient,
\begin{eqnarray}
&& a^{ijkl} = \rho\left( {\bar \theta}^2 g(z)  -2\bar \theta +1 \right)\delta^{ijkl}+\nonumber \\
&&+\rho \bar \theta \left(u^iu^j \delta^{kl} + u^i u^k \delta^{jl}+ u^i u^l \delta^{jk} + \right . \nonumber \\
&& \left . + u^k u^l\delta^{ij} + u^j u^l\delta^{ik} + u^j u^k \delta^{il} \right ) + \rho u^i u^j u^k u^l.
\end{eqnarray}
In power of the four Hermite coefficients we write their contractions to the Hermite polynomials themselves,

\begin{eqnarray}
&& aH=\rho\\
&& a^iH^i= \rho (\boldsymbol{\xi \cdot u})\\
&& a^{ij} H^{ij} = \rho \big[ \big( \bar \theta-1\big) (\boldsymbol{\xi}^2-D)+ (\boldsymbol{\xi \cdot u})^2 -\boldsymbol{u}^2\big]\nonumber \\
&& \\
&& a^{ijk}H^{ijk}=\rho \{ (\boldsymbol{\xi \cdot u})^3 - 3\boldsymbol{u}^2(\boldsymbol{\xi \cdot u}) + \big(\bar \theta-1\big)\cdot   \nonumber \\
&&\cdot \big[ 3(\boldsymbol{\xi \cdot u})\boldsymbol{\xi}^2 - 3(\boldsymbol{\xi \cdot u})D -6\big(\boldsymbol{\xi \cdot u}\big ) \big] \}\nonumber \\
&& \\
&& a^{ijkl}H^{ijkl} = \rho \{\left[ {\bar \theta}^2 g(z)  -2\bar \theta +1 \right] \cdot \nonumber \\
&& \cdot \big ( 3\boldsymbol{\xi}^4 - 6 D \boldsymbol{\xi}^2 -12\boldsymbol{\xi}^2 +3D^2 + 6D \big) +6\big( \bar \theta -1\big)\cdot \nonumber\\
&& \cdot \big [(\boldsymbol{\xi \cdot u})^2(\boldsymbol{\xi}^2-D-4) + \boldsymbol{u}^2 \big (D+2-\boldsymbol{\xi}^2 \big )\big ]+\nonumber \\
&& + \big [(\boldsymbol{\xi \cdot u})^4  - 6(\boldsymbol{\xi \cdot u})^2 \boldsymbol{u}^2 + 3 \boldsymbol{u}^4 \big ] \}.
\end{eqnarray}
From Eqs.(\ref{fhermite}) and (\ref{omega0}), one obtains that,
\begin{eqnarray}
&&\ f^{(0)}(\boldsymbol{x, \xi}) = \omega(\boldsymbol{\xi}) \left (a
H+ a^i H^i + \frac{1}{2!}a^{ij}H^{ij}+ \right.\nonumber \\
&&\left.+\frac{1}{3!}a^{ijk}H^{ijk}+\frac{1}{4!}a^{ijkl}H^{ijkl}\right).
\end{eqnarray}
Therefore the BE-FD function expanded in Hermite polynomials until the fourth order is,
\begin{eqnarray}\label{edf-cont}
&& f^{(0)}(\boldsymbol{x}, \boldsymbol{\xi})= \omega(\boldsymbol{\xi})\rho \Bigg\{ 1+ (\boldsymbol{\xi \cdot u} )\left( 1-\frac{1}{2}\boldsymbol{u}^2\right) + \nonumber \\
&&+ \frac{1}{6}(\boldsymbol{\xi \cdot u})^3 +(\boldsymbol{\xi \cdot u})^2 \left( \frac{1}{2} -\frac{1}{4}\boldsymbol{u}^2 \right)- \frac{1}{2}\boldsymbol{u}^2   + \frac{1}{8}\boldsymbol{u}^4+ \nonumber \\
&& +\frac{1}{24}(\boldsymbol{\xi \cdot u})^4 + \nonumber \\
&&+ \left( \bar \theta -1 \right)\cdot \bigg[\frac{1}{2}(\boldsymbol{\xi}^2-D)+ \frac{1}{2}(\boldsymbol{\xi \cdot u})(\boldsymbol{\xi}^2 - D-2) + \nonumber \\
&&+\frac{1}{4}(\boldsymbol{\xi \cdot u})^2 (\boldsymbol{\xi}^2-D-4) + \frac{1}{4}\boldsymbol{u}^2(D+2-\boldsymbol{\xi}^2)\bigg] \nonumber \\
&&+ \frac{1}{8}  \left[ {\bar \theta}^2 g(z)  -2\bar \theta +1 \right]  [\boldsymbol{\xi}^4+(D+2)(D-2\boldsymbol{\xi}^2)]\Bigg\},
\end{eqnarray}
where the definition of $\rho$, $\bar \theta$, and $g(z)$, are given in Eqs.(\ref{rhodef}), (\ref{bartheta}) and (\ref{g(z)}), respectively. In the appendix \ref{appendix-Taylor} we Taylor expand the MB distribution function thus providing an independent check that BE-FD and MB coincide in the limit that $g_{\nu}(z) \rightarrow z$.
\section{Gauss-Hermite quadrature}\label{quadrature}
We review the Gauss-Hermite quadrature, which is a way to obtain the values of integrals through discrete sums. In this way we transform our equilibrium distribution functions obtained in the previous section in a LBM scheme to numerically solve the BGK-Boltzmann equation. The Gauss-Hermite quadrature preserves the orthogonality of the Hermite
polynomial tensors in the Hilbert space~\cite{PhysRevE.73.056702}. Then the gaussian integrals can be
performed in a D-dimensional {\it discrete} space where the
microscopic velocity only takes the fixed  set of values,
$\boldsymbol{\xi}_{\alpha}$, $\alpha= 0, 1,\cdots,M_{\alpha}-1$, provided
that we introduce the set of weights $w_{\alpha}$. Essentially this
means that the gaussian integral over an arbitrary function $G \left
(\boldsymbol{\xi}
 \right )$ can be performed as a sum over a set of $M_{\alpha}$
 velocities:
\begin{eqnarray} \label{soma0}
\int d^D \boldsymbol{\xi}\, \omega(\boldsymbol{\xi}) G \left (\boldsymbol{\xi}
\right ) = \sum \limits_{\alpha} w_{\alpha} G \left (
\boldsymbol{\xi}_{\alpha}\right )
\end{eqnarray}
Thus  the orthonormality condition of the Hermite polynomials, shown
in Eq.(\ref{hermiteortho}), also holds in this discrete space:
\begin{eqnarray} \label{hermiteortho1}
&& \sum\limits_{\alpha} w_{\alpha}
H^{i_1i_2...i_N}(\boldsymbol{\xi}_{\alpha})
H^{j_1j_2...j_M}(\boldsymbol{\xi}_{\alpha}) = \\
&&= \delta^{N M} ( \delta^{i_1j_1}\delta^{i_2j_2}...\delta^{i_Nj_N}
+ \mbox{all permutations of j's}) \nonumber
\end{eqnarray}
Notice the two distinct integers, the truncation order of the
Hermite expansion of the equilibrium distribution, $N$, and also
$M$, the number of relations that the discrete set of velocities and
weights $w_{\alpha}$ must satisfy~\cite{PhysRevE.73.056702}. For
instance, we show below the $M=8$ relations that replace the
continuum Eqs. (\ref{int0})-(\ref{int6}).
\begin{eqnarray}
\sum\limits_{\alpha} w_{\alpha}=1
\label{sum0}
\end{eqnarray}
\begin{eqnarray}
\sum\limits_{\alpha} w_{\alpha} \xi^i_{\alpha} =0 \label{sum1}
\end{eqnarray}
\begin{eqnarray}
\sum\limits_{\alpha} w_{\alpha} \xi^i_{\alpha} \xi^j_{\alpha}= \delta^{i j} \label{sum2}
\end{eqnarray}
\begin{eqnarray}
\sum\limits_{\alpha} w_{\alpha} \xi^i_{\alpha} \xi^j_{\alpha} \xi^k_{\alpha}= 0 \label{sum3}
\end{eqnarray}
\begin{eqnarray}
\sum\limits_{\alpha} w_{\alpha}
\xi^i_{\alpha}\xi^j_{\alpha}\xi^k_{\alpha}\xi^l_{\alpha}
= \delta^{ijkl}\label{sum4}
\end{eqnarray}
\begin{eqnarray}
\sum\limits_{\alpha} w_{\alpha} \xi^i_{\alpha}\xi^j_{\alpha}\xi^k_{\alpha}\xi^l_{\alpha}\xi^m_{\alpha}= 0 \label{sum5}
\end{eqnarray}
\begin{eqnarray}
\sum\limits_{\alpha} w_{\alpha}
\xi^i_{\alpha}\xi^j_{\alpha}\xi^k_{\alpha}\xi^l_{\alpha}\xi^m_{\alpha}
\xi^n_{\alpha}= \delta^{ijklmn}, \label{sum6}
\end{eqnarray}
where the tensors $\delta^{i_1\cdots i_N}$ have been previously
defined. In this case the equilibrium distribution function of Eq.(\ref{edf-cont}) becomes
\begin{eqnarray} \label{edf-dis}
&& f^{(0)}_{\alpha}= w_{\alpha}\rho \Bigg\{ 1+ (\boldsymbol{\xi}_{\alpha}\boldsymbol{\cdot u} )\left( 1-\frac{1}{2}\boldsymbol{u}^2\right) + \nonumber \\
&&+ \frac{1}{6}(\boldsymbol{\xi}_{\alpha}\boldsymbol{\cdot u})^3 +(\boldsymbol{\xi}_{\alpha}\boldsymbol{\cdot u})^2 \left( \frac{1}{2} -\frac{1}{4}\boldsymbol{u}^2 \right)- \frac{1}{2}\boldsymbol{u}^2   + \frac{1}{8}\boldsymbol{u}^4+ \nonumber \\
&& +\frac{1}{24}(\boldsymbol{\xi}_{\alpha}\boldsymbol{\cdot u})^4 +\left( \bar \theta  -1 \right) \cdot \nonumber \\ &&\cdot \bigg[\frac{1}{2}(\boldsymbol{\xi}_{\alpha}^2-D)+ \frac{1}{2}(\boldsymbol{\xi}_{\alpha}\boldsymbol{\cdot u})(\boldsymbol{\xi}_{\alpha}^2 - D-2) + \nonumber \\
&&+\frac{1}{4}(\boldsymbol{\xi}_{\alpha}\boldsymbol{\cdot u})^2 (\boldsymbol{\xi}_{\alpha}^2-D-4) + \frac{1}{4}\boldsymbol{u}^2(D+2-\boldsymbol{\xi}_{\alpha}^2)\bigg] \nonumber \\
&&+ \frac{1}{8} \left[ {\bar \theta}^2 g(z)  -2\bar \theta +1 \right]  \cdot \nonumber \\ && \cdot [\boldsymbol{\xi}_{\alpha}^4+(D+2)(D-2\boldsymbol{\xi}_{\alpha}^2)]\Bigg\}.
\end{eqnarray}
Thus the continuum and discrete distribution functions basically
differ by the replacement of the gaussian function
$\omega(\boldsymbol{\xi})$ by the weights $w_{\alpha}$.

\section{Multi-scale expansion}\label{chapman-enskog}
In this section we perform the Chapman-Enskog analysis of the $N=4$
order BGK-Boltzmann theory and obtain the macroscopic hydrodynamical
equations. The local density, macroscopic velocity and energy are
obtained at any time and at any grid point through the definitions:
\begin{eqnarray}
& &\rho(\boldsymbol{x},t)=  \sum\limits_{\alpha}  f_{\alpha}(\boldsymbol{x}, t) \label{fneq1}\\
& &\boldsymbol{u}(\boldsymbol{x},t)
=\frac{1}{\rho(\boldsymbol{x},t)}\sum\limits_{\alpha} \boldsymbol{\xi} f_{\alpha}(\boldsymbol{x}, t) \label{fneq2}\\
& &\frac{D}{2} \bar \theta(\boldsymbol{x,t})
=\frac{1}{\rho(\boldsymbol{x},t)}\sum\limits_{\alpha}
\frac{\left[\boldsymbol{\xi}_{\alpha}-\boldsymbol{u}(\boldsymbol{x})\right]^2}{2}
f_{\alpha}(\boldsymbol{x}, t).\label{fneq3}
\end{eqnarray}
The last expression truly defines $\bar \theta$ and can be expressed as,
\begin{eqnarray}
\frac{D}{2} \bar \theta(\boldsymbol{x,t})+\frac{1}{2}\boldsymbol{u}(\boldsymbol{x},t)^2
=\frac{1}{\rho(\boldsymbol{x},t)}\sum\limits_{\alpha}\frac{\boldsymbol{\xi}_{\alpha}^2}{2}
f_{\alpha}(\boldsymbol{x}, t).\label{fneq4}
\end{eqnarray}
The so-called Chapman-Enskog relations must hold to assure the
macroscopic thermodynamic equations~\cite{succi,wolf-gladrow}:
\begin{eqnarray}
&&\sum_\alpha f_\alpha= \sum_\alpha f_\alpha^{(0)} \label{ch-ensk1}\\
&&\sum_\alpha f_\alpha \boldsymbol{\xi_\alpha} = \sum_\alpha
f_\alpha^{(0)} \boldsymbol{\xi_\alpha} \label{ch-ensk2}\\
&&\sum_\alpha f_\alpha \boldsymbol{\xi}_\alpha^2 = \sum_\alpha
f_\alpha^{(0)} \boldsymbol{\xi}_\alpha^2. \label{ch-ensk3}
\end{eqnarray}
These equations imply that $\rho= \sum\limits_{\alpha}  f_{\alpha}^{(0)}$, $\rho\boldsymbol{u}=\sum\limits_{\alpha} \boldsymbol{\xi} f_{\alpha}^{(0)}$, and $D\rho\bar \theta/2+\rho\boldsymbol{u}^2/2
=\sum\limits_{\alpha}\boldsymbol{\xi}_{\alpha}^2f_{\alpha}^{(0)}/2$.
The quantities $\rho$, $\boldsymbol{u}$, and $\theta$,
obtained from Eqs.(\ref{fneq1}), (\ref{fneq2}) and (\ref{fneq3}),
respectively, must be fed back into $f_\alpha^{(0)}$,
which, by its turn, sets the evolution to a new
$f_\alpha(\boldsymbol{x}, t)$, according to Eq.(\ref{disboltzeq0}). Next
we seek a formal solution of the distribution $f_\alpha$ in terms of
$f_\alpha^{(0)}$.  To obtain this solution  two key ingredients must
be considered. Firstly we Taylor expand $f_\alpha$ until second
order in the time step:
\begin{eqnarray}
&&f_\alpha(\boldsymbol{x}+ \boldsymbol{\xi_\alpha}\Delta t,t+\Delta
t)\cong f_\alpha(\boldsymbol{x},t)+\Delta t \left[\xi_\alpha^i \frac{\partial f_\alpha}{\partial x^i}+ \frac{\partial f_\alpha}{\partial t}\right]+\nonumber \\
&&+  \frac{(\Delta t)^2}{2} \left[  \frac{\partial^2
f_\alpha}{\partial t^2} + \xi_\alpha^i \xi_\alpha^j \frac{\partial^2
f_\alpha}{\partial x^i \partial x^j}  + 2 \xi^i_\alpha
\frac{\partial^2 f_\alpha}{\partial x^i\partial t } \right].
\end{eqnarray}
Then the discrete Boltzmann equation, Eq.(\ref{disboltzeq0}),
becomes in this approximation,
\begin{eqnarray}\label{boltzappro}
&&\xi_\alpha^i \frac{\partial f_\alpha}{\partial x^i}+  \frac{\partial f_\alpha}{\partial t}+ \frac{\Delta t}{2} \left[  \frac{\partial^2 f_\alpha}{\partial t^2} +  \xi_\alpha^i \xi_\alpha^j \frac{\partial^2 f_\alpha}{\partial x^i \partial x^j}  + 2 \xi^i_\alpha \frac{\partial^2 f_\alpha}{\partial x^i\partial t } \right] \cong \nonumber \\
&& \cong  - \frac{1}{\tau} \left( f_\alpha  - f^{(0)}_\alpha \right).
\label{LBM-Taylor}
\end{eqnarray}
Secondly we introduce the Chapman-Enskog expansion, which means to
expand the distribution function in terms of a parameter $\epsilon$
that represents the Knudsen's number.
\begin{equation}
f_\alpha = f_\alpha^{(0)} + \epsilon f_\alpha^{(1)}+ \epsilon^2
f_\alpha^{(2)}+ ... \label{dfunction_expansion}
\end{equation}
Notice that because of the Chapman-Enskog relations,
Eqs.(\ref{ch-ensk1}), (\ref{ch-ensk2}) and (\ref{ch-ensk3}), the
$\epsilon$ dependent terms must satisfy the following relations:
\begin{eqnarray}
&&\sum_\alpha f_\alpha^{(1)}=\sum_\alpha f_\alpha^{(2)}=0,\label{chens1}\\
&&\sum_\alpha \boldsymbol{\xi}_\alpha f_\alpha^{(1)}=\sum_\alpha \boldsymbol{\xi}_\alpha f_\alpha^{(2)}=0, \label{chens2}\\
&&\sum_\alpha\boldsymbol{\xi}_\alpha^2
f_\alpha^{(1)}=\sum_\alpha\boldsymbol{\xi}_\alpha^2 f_\alpha^{(2)}=0.\label{chens3}
\end{eqnarray}
The parameter $\epsilon$ also sets the scale for the derivatives of
time and space that are expanded in Eq.(\ref{dfunction_expansion}):
\begin{eqnarray}
&&\frac{\partial}{\partial t} = \epsilon \frac{\partial}{\partial
t_1} + \epsilon^2 \frac{\partial}{\partial t_2},
\label{t_deriv_expansion} \\
&&\frac{\partial}{\partial x^i} = \epsilon \frac{\partial}{\partial
x^i_1}. \label{x_deriv_expansion}
\end{eqnarray}
We stress for later purposes that the cross time and position derivative is given by,
\begin{eqnarray}\label{tx_deriv_expansion}
\frac{\partial}{\partial x^i}\frac{\partial}{\partial t} = \epsilon^2 \frac{\partial}{\partial t_1} \frac{\partial}{\partial x_1^i},
\end{eqnarray}
to the studied  order $\epsilon^2$.
Applying (\ref{dfunction_expansion}), (\ref{t_deriv_expansion}) and
(\ref{x_deriv_expansion}) in (\ref{LBM-Taylor}),
\begin{eqnarray}
&&\epsilon \left[ \xi^i_\alpha \frac{\partial
f^{(0)}_\alpha}{\partial x^i_1} + \frac{\partial f^{(0)}_\alpha}{\partial
t_1} \right] +\epsilon^2 \left[ \xi^i_\alpha \frac{\partial
f_\alpha^{(1)}}{\partial x^i_1} + \frac{\partial
f_\alpha^{(1)}}{\partial t_1} + \frac{\partial
f_\alpha^{(0)}}{\partial t_2}+ \right.\nonumber \\
&&\left.+\frac{\Delta t}{2} \frac{\partial^2
f_\alpha^{(0)}}{\partial t_1^2} + \frac{\Delta t}{2} \xi^i_\alpha
\xi^j_\alpha \frac{\partial ^2 f_\alpha ^{(0)}}{\partial x^i_1
\partial x^j_1} + \Delta t \xi^i_\alpha \frac{\partial^2
f_\alpha^{(0)}}{\partial x^i_1 \partial t_1} \right] = \nonumber \\
&& -\frac{1}{\tau} [ \epsilon f_\alpha^{(1)} + \epsilon^2 f^{(2)}
_\alpha + ... ].
\end{eqnarray}
Next we collect the terms of same order, and the first order terms
give that,
\begin{equation}
- \frac{1}{\tau}  f_\alpha^{(1)}=  \left[ \xi_\alpha^i
\frac{\partial f_\alpha^{(0)}}{\partial x^i_1} + \frac{\partial
f_\alpha^{(0)}}{\partial t_1} \right]. \label{f1_equation}
\end{equation}
For second order one obtains that,
\begin{eqnarray}
&& - \frac{1}{\tau} f_\alpha^{(2)}= \xi_\alpha^i \frac{\partial
f_\alpha^{(1)}}{\partial x^i_1}+ \frac{f_\alpha^{(1)}}{\partial t_1}
+ \frac{\partial f_\alpha^{(0)}}{\partial t_2} + \frac{\Delta t}{2}
\frac{\partial^2 f_\alpha^{(0)}}{\partial t_1^2} + \nonumber \\
&&+ \frac{\Delta t}{2} \xi_\alpha^i \xi_\alpha^j \frac{\partial^2
f_\alpha^{(0)}}{\partial x^i_1 \partial x^j_1}+ \Delta t
\xi_\alpha^i \frac{\partial^2 f_\alpha^{(0)}}{\partial x^i_1
\partial t_1}. \label{f2a_equation}
\end{eqnarray}
We derive Eq.(\ref{f1_equation}) with respect to $t_1$ and $x^i_1$ to find
that Eq.(\ref{f2a_equation}) becomes,
\begin{equation}
-\frac{1}{\tau}f_\alpha^{(2)}= \frac{\partial f_\alpha
^{(0)}}{\partial t_2} + \left( 1- \frac{\Delta t}{2 \tau} \right)
\left[ \frac{\partial f_\alpha^{(1)}}{\partial t_1} + \xi_\alpha^i
\frac{\partial f_\alpha^{(1)}}{\partial x^i_1} \right].
\label{f2_equation}
\end{equation}
In summary we found a solution for the discrete Boltzmann
(Eq.(\ref{boltzappro})), which is the distribution $f_\alpha$, given
by Eq.(\ref{dfunction_expansion}), as a function of $f_\alpha^{(0)}$
and its derivatives through Eqs.(\ref{f1_equation}) and
(\ref{f2_equation}). It remains to reconstruct the time and the
position, which are defined by Eqs.(\ref{t_deriv_expansion}) and
(\ref{x_deriv_expansion}). This will be done separately for the
mass, momentum and energy.

\subsection{Conservation of mass}
To obtain the continuity equation, one must sum over all directions $\alpha$ in Eq.(\ref{f1_equation}),
\begin{equation}
\frac{1}{\tau} \sum_\alpha f_\alpha^{(1)}=-\frac{\partial}{\partial
t_1} \sum_\alpha f_\alpha^{(0)} - \frac{\partial}{\partial x^i_1}
\sum_\alpha f_\alpha^{0} \xi^i_\alpha.
\end{equation}
Recall the definitions of $\rho$, $\boldsymbol{u}$,
Eqs.(\ref{fneq1}) and (\ref{fneq2}), and the Chapman-Enskog
relations of Eqs.(\ref{ch-ensk1}) and (\ref{ch-ensk2}):
\begin{equation}
\frac{\partial \rho}{\partial t_1} + \frac{\partial}{\partial
x^i_1}(\rho u^i) =0.
\end{equation}
This is not yet the continuity equation as the time
derivative is over $t_1$ instead of $t$. To fix it,
sum over $\alpha$ in Eq.(\ref{f2_equation}):
\begin{eqnarray}
&& -\frac{1}{\tau} \sum_\alpha f_\alpha^{(2)} =
\frac{\partial}{\partial t_2} \sum_\alpha f_\alpha^{(0)} + \nonumber\\
&&+ \left( 1-\frac{\Delta t}{2 \tau} \right) \left[
\frac{\partial}{\partial t_1}\sum_\alpha f^{(1)}_\alpha +
\sum_\alpha \xi_\alpha^i \frac{\partial f_\alpha^{(1)}}{\partial
x^i_1} \right].
\end{eqnarray}
The above equation is no more than,
\begin{equation}
\frac{\partial \rho}{\partial t_2}=0.
\end{equation}
To reconstruct the time and space derivative, take
Eqs.(\ref{t_deriv_expansion}) and (\ref{x_deriv_expansion}) applied
to the density, respectively,
\begin{eqnarray}
&& \frac{\partial \rho}{\partial t}= \epsilon \frac{\partial
\rho}{\partial t_1}+ \epsilon^2 \frac{\partial \rho}{\partial t_2} =
\epsilon \frac{\partial \rho}{\partial t_1}\\
&&\frac{\partial \rho}{\partial x^i} = \epsilon \frac{\partial
\rho}{\partial x^i_1}.
\end{eqnarray}
Then the continuity equation, Eq.(\ref{continuity}) is obtained.

\subsection{Conservation of momentum}
The derivation of the conservation of momentum equation follows
similar steps, which means that
Eqs.(\ref{f1_equation}) and (\ref{f2_equation}) are multiplied by $\xi_\alpha^i$ and next summed over $\alpha$.
\begin{equation}
- \frac{1}{\tau} \sum_\alpha \xi^i_\alpha f^{(1)_\alpha}=
\frac{\partial}{\partial x^j_1}\sum_\alpha \xi^i_\alpha \xi^j_\alpha
f_\alpha^{(0)} + \frac{\partial}{\partial t_1}\sum_\alpha
\xi_\alpha^i f^{(0)}_\alpha.
\end{equation}
Using the Chapman-Enskog relation of Eq.(\ref{ch-ensk2}), it follows that,
\begin{equation}
\frac{\partial}{\partial t_1}\sum_\alpha
\xi_\alpha^i f^{(0)}_\alpha + \frac{\partial}{\partial
x^j_1} \sum_\alpha \xi_\alpha^i \xi^j_\alpha f^{(0)}_\alpha=0.
\label{f1_times_xi}
\end{equation}
Next from Eq.(\ref{f2_equation}) one obtains that,
\begin{eqnarray}
&& - \frac{1}{\tau} \sum_\alpha \xi_\alpha^i f^{(2)}_\alpha =
\frac{\partial}{\partial t_2} \sum_\alpha\xi_\alpha^i f_\alpha^{(0)}
+ \\
&&+\left(1-\frac{\Delta t}{2 \tau}\right) \left[
\frac{\partial}{\partial t_1} \sum_\alpha \xi_\alpha^i
f_\alpha^{(1)} + \frac{\partial}{\partial x^j_1} \sum_\alpha
\xi_\alpha^i \xi^j_\alpha f^{(1)}_\alpha \right].\nonumber
\end{eqnarray}
We apply Eq.(\ref{chens2}) twice in the above equation to obtain that,
\begin{eqnarray}
&&\frac{\partial}{\partial t_2} \sum_\alpha
\xi_\alpha^i f^{(0)}_\alpha +\left( 1- \frac{\Delta
t}{2 \tau} \right) \frac{\partial}{\partial x^j_1} \sum_\alpha
\xi_\alpha^i \xi_\alpha^j f^{(1)}_\alpha=0. \nonumber \\
&& \label{f2_times_xi}
\end{eqnarray}
Next consider the construction $\epsilon$ times Eq.(\ref{f1_times_xi}) plus $\epsilon ^2$ times Eq.(\ref{f2_times_xi}):
\begin{eqnarray}
&& \left( \epsilon \frac{\partial}{\partial t_1} + \epsilon^2
\frac{\partial}{\partial t_2} \right)\sum_\alpha
\xi_\alpha^i f^{(0)}_\alpha + \epsilon
\frac{\partial}{\partial x^j_1} \sum_\alpha \xi_\alpha^i
\xi_\alpha^j f^{(0)}_\alpha + \nonumber\\
&&+ \epsilon ^2 \left(1-\frac{\Delta t }{2 \tau} \right)
\frac{\partial}{\partial x^j_1} \sum_\alpha
\xi_\alpha^i \xi_\alpha^j f^{(1)}_\alpha=0.
\end{eqnarray}
We use Eqs.(\ref{fneq2}), (\ref{ch-ensk2}) and introduce
Eq.(\ref{f1_equation}) into the above equation to write it solely in
terms of moments of $f_\alpha^{(0)}$:
\begin{eqnarray}
&&\left( \epsilon \frac{\partial}{\partial t_1} + \epsilon^2 \frac{\partial}{\partial t_2}\right) (\rho u^i) + \epsilon \frac{\partial}{\partial x_1^j} \sum_\alpha \xi_\alpha^i \xi_\alpha^j f_\alpha^{(0)}- \nonumber \\
&&- \tau \epsilon^2 \left( 1-\frac{\Delta t}{2\tau} \right) \frac{\partial}{\partial x_1^j} \frac{\partial}{\partial x^k_1} \sum_\alpha \xi_\alpha^i \xi_\alpha^j \xi_\alpha^k f_\alpha^{(0)}- \nonumber \\
&&-\tau \epsilon^2\left(1-\frac{\Delta t}{2\tau}\right) \frac{\partial}{\partial x_1^j} \frac{\partial}{\partial t_1}\sum_\alpha \xi_\alpha^i \xi_\alpha^j f^{(0)}_\alpha =0.
\end{eqnarray}
The true time and position derivatives are given by Eqs.(\ref{t_deriv_expansion}), (\ref{x_deriv_expansion}) and (\ref{tx_deriv_expansion}), and the  above equation acquires the form,
\begin{eqnarray}
&& \frac{\partial}{\partial t}(\rho u^i) + \frac{\partial}{\partial x^j} \pi^{ij} - \tau \left( 1-\frac{\Delta t}{2\tau} \right) \frac{\partial}{\partial x^j}\frac{\partial}{\partial x^k} \pi^{ijk} - \noindent \\
&&-\tau \left( 1-\frac{\Delta t}{2\tau} \right) \frac{\partial}{\partial x^i} \frac{\partial}{\partial t} \pi^{ij}=0,
\label{master_eq_momentum}
\end{eqnarray}
by definition of $\pi^{ij} \equiv \sum_\alpha \xi_\alpha^i \xi_\alpha^j f^{(0)}_\alpha$, $\pi^{ijk} \equiv \sum_\alpha \xi_\alpha^i \xi_\alpha^j \xi_\alpha^k f^{(0)}_\alpha$. We call it the master equation for the conservation of momentum. To determine the tensors $\pi^{ij}$ and $\pi^{ijk}$ in terms of the macroscopic quantities  one must introduce $f_\alpha^{(0)}$ given by Eq.(\ref{edf-dis}).
To accomplish such task one must invoke some relations calculated in appendix \ref{usefulrelations}. These are Eqs.(\ref{ur1}), (\ref{ur2}), (\ref{ur3}), (\ref{ur4}) and (\ref{ur5}) to calculate $\pi^{ij}$ and Eqs.(\ref{ur6}), (\ref{ur7}) and (\ref{ur8}) to derive $\pi^{ijk}$. After some algebra we find that,
\begin{eqnarray}\label{tensor-pi}
&&\pi^{ij}=\rho\left( \bar \theta \delta^{ij}+ u^i u^j \right)\\
&&\pi^{ijk}= \rho \left[ \bar \theta\left(u^k\delta^{ij}+u^j\delta^{ik}+u^i\delta^{jk}\right) + u^i u^j u^k \right].\nonumber
\end{eqnarray}
Substituting these results in equation (\ref{master_eq_momentum}),
we find that the equation for the momentum conservation is,
\begin{eqnarray}
&&\frac{\partial}{\partial t }\left(\rho u^i\right) + \frac{\partial}{\partial x^j} \left[ \rho\left( \bar \theta \delta^{ij} + u^i u^j \right) \right] - \tau\left(1-\frac{\Delta t}{2\tau}\right)\cdot \nonumber \\
&&\cdot \frac{\partial}{\partial x^j}\frac{\partial}{\partial x^k} \left\{ \rho\left[\bar \theta \left(u^k\delta^{ij} + u^j \delta^{ik}+ u^i \delta^{jk} \right) + u^i u^j u^k \right]\right\} - \nonumber \\
 &&-\tau \left(1-\frac{\Delta t}{2\tau} \right) \frac{\partial}{\partial x^i} \frac{\partial}{\partial t} \left[ \rho \left( \bar \theta \delta^{ij} + u^i u^j \right)\right]=0. \label{eq_momentum_with_crossing_derivative}
\end{eqnarray}
Notice that in the above equation the temperature is only present
through the pseudo temperature. The last step is to replace the
cross derivative term $\partial^2/\partial t \partial x^k$ by a
positional derivative term, $\partial^2/\partial x^i \partial x^j$ .
This task is carried in the appendix~\ref{appendix-cross-derivative}
and leads to the momentum equation of Eq.(\ref{navierstokeseq}).

\subsection{Conservation of energy}
The derivation of the conservation of energy equation also follows
from Eqs.(\ref{f1_equation}) and (\ref{f2_equation}), but in this
case multiplied by $\boldsymbol{\xi}_\alpha^2$ instead, and summed
over $\alpha$.
\begin{equation}
-\frac{1}{\tau} \sum_\alpha f_\alpha^{(1)} \boldsymbol{\xi}_\alpha^2
= \sum_\alpha \boldsymbol{\xi}_\alpha^2 \xi_\alpha^j \frac{\partial
f_\alpha^{(0)}}{\partial x^j_1} + \sum_\alpha \boldsymbol{\xi}_\alpha^2 \frac{\partial f_\alpha^{(0)}}{\partial t_1}.
\end{equation}
The Chapman-Enskog relation of Eq.(\ref{ch-ensk3}) renders the left
side of the above equation equal to zero, and so,
\begin{equation}
\frac{\partial}{\partial
t_1} \sum_\alpha \boldsymbol{\xi}_\alpha^2 f_\alpha^{(0)}+
\frac{\partial}{\partial x^j_1} \sum_\alpha \boldsymbol{\xi}_\alpha^2 \xi^j_\alpha f_\alpha^{(0)}  =0.
\label{f1_times_xi_xi}
\end{equation}
Similarly Eq.(\ref{f2_equation}) leads to the following equation.
\begin{eqnarray}
&&-\frac{1}{\tau} \sum_\alpha \boldsymbol{\xi}_\alpha^2
f_\alpha^{(2)} = \frac{\partial}{\partial t_2} \sum_\alpha
\boldsymbol{\xi}_\alpha^2 f_\alpha^{(0)} + \nonumber \\
&&+\left( 1-\frac{\Delta t}{2\tau} \right) \frac{\partial}{\partial
t_1} \sum_\alpha \boldsymbol{\xi}_\alpha^2  f_\alpha^{(1)} +
\nonumber \\
&&+\left( 1-\frac{\Delta t}{2\tau} \right) \frac{\partial}{\partial
x^j_1} \sum_\alpha \boldsymbol{\xi}_\alpha^2 \xi_\alpha^j
f_\alpha^{(1)}.
\end{eqnarray}
The Chapman-Enskog relation of Eq.(\ref{ch-ensk3}) renders two terms
null, including the left side.
\begin{equation}
\frac{\partial}{\partial t_2} \sum_\alpha \boldsymbol{\xi}_\alpha^2
f_\alpha^{(0)} + \left( 1-\frac{\Delta t}{2\tau} \right)
\frac{\partial}{\partial x^j_1} \sum_\alpha \boldsymbol{\xi}_\alpha^2 \xi_\alpha^j f_\alpha^{(1)}=0. \label{f2_times_xi_xi}
\end{equation}
To get the time and the position derivatives consider
Eqs.(\ref{t_deriv_expansion}) and (\ref{x_deriv_expansion}), and so, take
Eq.(\ref{f1_times_xi_xi}) times $\epsilon$ plus Eq.
(\ref{f2_times_xi_xi}) times $\epsilon^2$.
\begin{eqnarray}
&&\left( \epsilon \frac{\partial}{\partial t_1} + \epsilon^2 \frac{\partial}{\partial t_2}\right) \sum_\alpha \boldsymbol{\xi}_\alpha^2
f_\alpha^{(0)} + \epsilon \frac{\partial}{\partial x_1^j} \sum_\alpha \boldsymbol{\xi}_\alpha^2 \xi_\alpha^j f_\alpha^{(0)}+ \nonumber \\
&&+\epsilon^2 \left( 1-\frac{\Delta t}{2\tau} \right) \frac{\partial}{\partial x_1^j} \sum_\alpha \boldsymbol{\xi}_\alpha^2\xi_\alpha^j  f_\alpha^{(1)} =0.
\end{eqnarray}
Introduce Eq.(\ref{fneq4}) and Eq.(\ref{f1_equation}) to obtain that,
\begin{eqnarray}
&&\left( \epsilon \frac{\partial}{\partial t_1} + \epsilon^2 \frac{\partial}{\partial t_2}\right) \sum_\alpha \boldsymbol{\xi}_\alpha^2
f_\alpha^{(0)}  + \epsilon \frac{\partial}{\partial x_1^j} \sum_\alpha \boldsymbol{\xi}_\alpha^2 \xi_\alpha^j f_\alpha^{(0)}- \nonumber \\
&&- \tau \epsilon^2 \left( 1-\frac{\Delta t}{2\tau} \right) \frac{\partial}{\partial x_1^j} \frac{\partial}{\partial x^k_1} \sum_\alpha \boldsymbol{\xi}_\alpha^2\xi_\alpha^j \xi_\alpha^k f_\alpha^{(0)}- \nonumber \\
&&-\tau \epsilon^2\left(1-\frac{\Delta t}{2\tau}\right) \frac{\partial}{\partial x_1^j} \frac{\partial}{\partial t_1}\sum_\alpha \boldsymbol{\xi}_\alpha^2\xi_\alpha^j f^{(0)}_\alpha =0.
\end{eqnarray}
Then one gets,
\begin{eqnarray}
&&\frac{\partial}{\partial t}\phi + \frac{\partial}{\partial
x^j}\phi^j -\\
&& -\tau \left( 1-\frac{\Delta t}{2 \tau} \right)
\frac{\partial}{\partial x^j}\frac{\partial}{\partial x^k}\phi^{jk} -\tau \left( 1-\frac{\Delta t}{2 \tau} \right)
\frac{\partial}{\partial x^j}\frac{\partial}{\partial t}\phi^{j} =0. \nonumber
\label{energy_eq}
\end{eqnarray}
This is the master equation for the conservation of energy. We
define $\phi\equiv \sum_\alpha \boldsymbol{\xi_\alpha}^2
f_\alpha^{(0)}/2$, $\phi^j \equiv \sum_\alpha
\boldsymbol{\xi_\alpha}^2 \xi_\alpha^j f_\alpha^{(0)}/2$ and
$\phi^{jk} \equiv \sum_\alpha \boldsymbol{\xi_\alpha}^2 \xi_\alpha^j
\xi_\alpha^k f_\alpha^{(0)}/2$ and use
Eq.(\ref{tx_deriv_expansion}). It is straightforward to calculate
$\phi$, $\phi^j$ and $\phi^{jk}$ from the expressions developed in
the appendix. Then one determines these tensors in terms of the
macroscopic quantities contained in $f_\alpha^{(0)}$, given by
Eq.(\ref{edf-dis}).
\begin{eqnarray}
&&\phi = \frac{1}{2}\rho\boldsymbol{u}^2 + \frac{D}{2}\rho \bar \theta \label{phi}\\
&&\phi^j = \frac{\rho}{2}\left[ u^j \boldsymbol{u}^2 + \bar \theta \left (D+2 \right ) u^j \right]\label{phij}\\
&&\phi^{jk} = \frac{\rho}{2}\left[ \boldsymbol{u}^2 u^j u^k + \bar \theta \left (D+4 \right ) u^j u^k+\right . \nonumber \\
&& \left . +\bar \theta  \boldsymbol{u}^2 \delta^{jk} + {\bar\theta}^2g(z)\left(D+2\right)\delta^{jk} \right].\label{phijk}
\end{eqnarray}
Notice that among the above three tensors only $\phi^{jk}$ contains
an explicit quadratic term in the temperature, which leads to the
presence of the $g(z)$ function in Eq.(\ref{g(z)}). Thus the energy
equation cannot be purely expressed in terms of $\rho$,
$\boldsymbol{u}$ and $\bar \theta$. We write it with a $g(z)-1$ term
intentionally in its right side. The left side corresponds to the
classical (MB) case, replacing $\bar \theta$ by $\theta$,
\begin{eqnarray}
&&\frac{\partial}{\partial t} \left( \frac{\rho}{2}\boldsymbol{u}^2 + \frac{\rho}{2} D\bar \theta  \right) + \frac{\partial}{\partial x^j} \left[ \left( \frac{\rho}{2}\boldsymbol{u}^2 + \frac{\rho}{2}\bar \theta  (D+2)\right) u^j\right]- \nonumber \\
&&- \tau\left( 1-\frac{\Delta t}{2\tau} \right) \frac{\partial}{\partial x^j} \frac{\partial}{\partial x^k} \left[ \frac{\rho}{2}\boldsymbol{u}^2 u^j u^k+ \right . \nonumber \\
&&\left . + \frac{\rho}{2}\bar \theta (D+4) u^j u^k + \frac{\rho}{2}\bar \theta  \boldsymbol{u}^2 \delta^{jk} + \frac{\rho}{2}{\bar\theta}^2 (D+2)\delta^{jk} \right] -\nonumber \\
&&- \tau \left( 1-\frac{\Delta t}{2 \tau} \right)\frac{\partial}{\partial x^j} \frac{\partial}{\partial t}\left[ \left(\frac{\rho}{2} \boldsymbol{u}^2 + \frac{\rho}{2} \bar \theta  (D+2)\right) u^j \right] = \nonumber \\
&&=\tau\left( 1-\frac{\Delta t}{2\tau} \right) \frac{\partial}{\partial x^j} \frac{\partial}{\partial x^k} \left[  \frac{\rho}{2}{\bar\theta}^2 \left(g(z)-1\right) (D+2)\delta^{jk} \right]. \nonumber \\
\label{macro-e2}
\end{eqnarray}
Similarly to the momentum case, the cross derivative term $\partial^2/\partial t \partial x^k$ is replaced by a positional derivative term, $\partial^2/\partial x^i \partial x^j$, in the appendix~\ref{appendix-cross-derivative}, and this leads to Eq.(\ref{energy-3}).

\section{The Hermite polynomial expansion to order N=3}\label{n=3}
Here we show that the Hermite polynomial expansion of the BE-FD
equilibrium distribution function must be carried until $N=4$ order
to obtain meaningful results. The $N=3$ order simply does not
correctly describe the energy equation. The BE-FD function
equilibrium distribution function expanded until order $N=3$ is
given by,
\begin{eqnarray}\label{edf-cont3}
&& {\hat f}^{(0)}_{\alpha}= w_{\alpha}\rho \Bigg\{ 1- \frac{1}{2}\boldsymbol{u}^2   + (\boldsymbol{\xi}_{\alpha} \boldsymbol{\cdot u} )\left( 1-\frac{1}{2}\boldsymbol{u}^2\right) + \nonumber \\
&&+\frac{1}{2}(\boldsymbol{\xi}_{\alpha}\boldsymbol{\cdot u})^2+ \frac{1}{6}(\boldsymbol{\xi}_{\alpha} \boldsymbol{\cdot u})^3 +  \\
&&+ \left( \bar \theta -1 \right)\cdot \bigg[\frac{1}{2}(\boldsymbol{\xi}_{\alpha}^2-D)+ \frac{1}{2}(\boldsymbol{\xi}_{\alpha}\boldsymbol{\cdot u})(\boldsymbol{\xi}_{\alpha}^2 - D-2) \bigg]\Bigg\}. \nonumber
\end{eqnarray}
Notice that for the above equilibrium distribution function the
fugacity $z$ is only present through the pseudo temperature $\bar
\theta$ of Eq.(\ref{bartheta}), but this is not so for the
equilibrium distribution function of Eq.(\ref{edf-cont}). This shows
that the quantum and the classical macroscopic hydrodynamical
equations are identical until $N=3$ order by mapping the pseudo
variables into their counterparts. In order $N=4$ this mapping holds
for the mass and momentum but not for the energy equation. This
conclusion can be reached by analysing the $N=3$ order construction
of the momentum and energy tensors, given by Eqs.(\ref{tensor-pi}),
(\ref{phi}), (\ref{phij}) and (\ref{phijk}). Interestingly, the
momentum tensors of Eq.(\ref{tensor-pi}) are the same in order $N=3$
and $N=4$, as seen below.
\begin{eqnarray}
&&\sum_\alpha \xi_\alpha^i  {\hat f}^{(0)}_\alpha=\rho u^i \\
&&\sum_\alpha \xi_\alpha^i \xi_\alpha^j {\hat f}^{(0)}_\alpha=\rho\left( \bar \theta \delta^{ij}+ u^i u^j \right)\\
&&\sum_\alpha \xi_\alpha^i \xi_\alpha^j \xi_\alpha^k {\hat f}^{(0)}_\alpha= \rho \left[ \bar \theta\left(u^k\delta^{ij}+u^j\delta^{ik}+u^i\delta^{jk}\right) + u^i u^j u^k \right].\nonumber \\
\end{eqnarray}
Thus we conclude that to obtain Eq.(\ref{master_eq_momentum}) and
Eq.(\ref{eq_momentum_with_crossing_derivative}) it is just enough to
go in the expansion to  order $N=3$. Consequently the quantum Navier
Stokes equation, namely, Eq.(\ref{navierstokeseq}), is obtained in
order $N=3$ and higher order terms, such as $N=4$ terms, do not
contribute to it. Thus concerning the momentum equation, classical
and quantum fluids have formally the same macroscopic description.
However the same does not hold for the energy equation, as we learn
by the expressions of the energy tensors in $N=3$ order. They are
given by,
\begin{eqnarray}
&&\frac{1}{2}\sum_\alpha \boldsymbol{\xi_\alpha}^2 {\hat f}_\alpha^{(0)} = \frac{1}{2}\rho\boldsymbol{u}^2 + \frac{D}{2}\rho \bar \theta \\
&&\frac{1}{2}\sum_\alpha \boldsymbol{\xi_\alpha}^2 \xi_\alpha^j {\hat f}_\alpha^{(0)} = \frac{\rho}{2}\left[ u^j \boldsymbol{u}^2 + \bar \theta \left (D+2 \right ) u^j \right]\\
&&\frac{1}{2}\sum_\alpha \boldsymbol{\xi_\alpha}^2 \xi_\alpha^j \xi_\alpha^k {\hat f}_\alpha^{(0)} = \frac{\rho}{2}\left[\left (D+4 \right ) u^j u^k+\right . \nonumber \\
&& \left . + \boldsymbol{u}^2 \delta^{jk} + \left(2{\bar\theta}-1\right)\left(D+2\right)\delta^{jk} \right].
\end{eqnarray}
The last tensor, $\sum_\alpha \boldsymbol{\xi_\alpha}^2 \xi_\alpha^j \xi_\alpha^k {\hat f}_\alpha^{(0)}/2$, differs from its $N=4$ counterpart, as seen by comparison to Eq.(\ref{phijk}).
The first two other tensors are identical in $N=3$ and $N=4$ order, as seen by comparison to Eqs.(\ref{phi}) and (\ref{phij}).
Thus with the exception of a single energy tensor all other ones are identical in $N=3$ and $N=4$ order. Nevertheless this single tensor renders the $N=3$ energy equation unfit to describe the energy evolution of the quantum fluid. Hereafter we develop a heuristic procedure to obtain the $N=3$ tensor from its $N=4$ counterpart, given by Eq.(\ref{phijk}). Consider the following approximations in the $N=4$ tensor, $\sum_\alpha \boldsymbol{\xi_\alpha}^2 \xi_\alpha^j \xi_\alpha^k {\hat f}_\alpha^{(0)}/2 = \rho [ \boldsymbol{u}^2 u^j u^k + \bar \theta (D+4) u^j u^k+\bar \theta  \boldsymbol{u}^2 \delta^{jk} + {\bar\theta}^2g(z)(D+2)\delta^{jk} ]$. Firstly drop the fourth power terms in the velocity. Notice that $\bar \theta-1$ is of the order of velocity square, as shown in appendix~\ref{appendix-Taylor}, and so, terms multiplied by the second power in the velocity are of maximum order, and therefore, also dropped. Apply $g(z)=1$, as we know that such function does not exist in $N=3$ order because it is not contained in ${\hat f}^{(0)}_{\alpha}$. Therefore take the above tensor in the limit that $\bar \theta = (\bar \theta-1)+1\rightarrow 1$ and $\bar \theta^2=(\bar \theta-1)^2+2\bar \theta-1\rightarrow 2\bar \theta-1$ to obtain the above $N=3$  expression for this tensor. In conclusion, contributions of $N=4$ order to the last tensor, $\sum_\alpha \boldsymbol{\xi_\alpha}^2 \xi_\alpha^j \xi_\alpha^k {\hat f}_\alpha^{(0)}/2$, are important and necessary to reach the energy equation of Eq.(\ref{macro-e2}). Hence neither the energy of Eq.(\ref{energy-3}) nor the thermal coefficient of Eq.(\ref{thermal_coeff}) can be obtained in $N=3$ order. This can only be done in order $N=4$, as shown in this paper.

\section{The dilute quantum fluid}\label{dilute}
We develop here the approximation of the dilute quantum (BE-FD)
system, which is very near to the classical (MB) limit. This is the
limit of small fugacity, $z \ll 1$ taken in the function of
Eq.(\ref{function-g})~\cite{pathria}:
\begin{eqnarray}\label{gnu-approx}
g_{\nu}(z)= z \mp \frac{z^2}{2^{\nu}} + \cdots
\end{eqnarray}
where the assignment is (+) BE and (-) FD, respectively. We intend
to carry here just the first order corrections to the classical (MB)
fluid which are linear in the fugacity. Then we Taylor expand the
function $g(z)$ to obtain that,
\begin{eqnarray}\label{g-approx}
g(z)= 1 \mp \frac{z}{2^{\frac{D}{2}+2}} + \cdots
\end{eqnarray}
In this order the density and the pseudo temperature are given by,
\begin{eqnarray}\label{rho-theta-approx}
&& \rho= \left(2\pi\theta \right)^{\frac{D}{2}}z + \cdots \\
&& \bar \theta = \theta \left( 1 \pm \frac{z}{2^{\frac{D}{2}+2}} + \cdots\right).
\end{eqnarray}
Within this order we can write the fugacity and so the function
$g(z)$ in terms of the density and of the pseudo temperature,
\begin{eqnarray}\label{g-approx}
g \approx 1 \mp \frac{1}{2^{\frac{D}{2}+2}}\frac{\rho}{\left( 2\pi \bar\theta \right)^{\frac{D}{2}}}
\end{eqnarray}
under the assumption that
\begin{eqnarray}
&& z \approx \frac{\rho}{\left( 2\pi \bar\theta \right)^{\frac{D}{2}}} \ll 1.
\end{eqnarray}
Under this approximation the function $g$ can be brought back to the
BE-FD equilibrium distribution function of Eq.(\ref{edf-dis}) that
becomes a function of $\rho$, $\bar\theta$ and $\boldsymbol{u}$.

\begin{eqnarray} \label{edf-dis-dil}
&& f^{(0)}_{\alpha}= w_{\alpha}\rho \Bigg\{ 1+ (\boldsymbol{\xi}_{\alpha}\boldsymbol{\cdot u} )\left( 1-\frac{1}{2}\boldsymbol{u}^2\right) + \nonumber \\
&&+ \frac{1}{6}(\boldsymbol{\xi}_{\alpha}\boldsymbol{\cdot u})^3 +(\boldsymbol{\xi}_{\alpha}\boldsymbol{\cdot u})^2 \left( \frac{1}{2} -\frac{1}{4}\boldsymbol{u}^2 \right)- \frac{1}{2}\boldsymbol{u}^2   + \frac{1}{8}\boldsymbol{u}^4+ \nonumber \\
&& +\frac{1}{24}(\boldsymbol{\xi}_{\alpha}\boldsymbol{\cdot u})^4 +\left( \bar \theta  -1 \right) \cdot \nonumber \\ &&\cdot \bigg[\frac{1}{2}(\boldsymbol{\xi}_{\alpha}^2-D)+ \frac{1}{2}(\boldsymbol{\xi}_{\alpha}\boldsymbol{\cdot u})(\boldsymbol{\xi}_{\alpha}^2 - D-2) + \nonumber \\
&&+\frac{1}{4}(\boldsymbol{\xi}_{\alpha}\boldsymbol{\cdot u})^2 (\boldsymbol{\xi}_{\alpha}^2-D-4) + \frac{1}{4}\boldsymbol{u}^2(D+2-\boldsymbol{\xi}_{\alpha}^2)\bigg] \nonumber \\
&&+ \frac{1}{8} \left[ {\bar \theta}^2 \left( 1 \mp \frac{1}{2^{\frac{D}{2}+2}}\frac{\rho}{\left( 2\pi \bar\theta \right)^{\frac{D}{2}}}\right)  -2\bar \theta +1 \right]  \cdot \nonumber \\ && \cdot [\boldsymbol{\xi}_{\alpha}^4+(D+2)(D-2\boldsymbol{\xi}_{\alpha}^2)]\Bigg\}.
\end{eqnarray}
Thus the LBM for the dilute quantum fluid can be developed through
 the variables density $\rho$, macroscopic velocity $\boldsymbol{u}$
and pseudo temperature $\bar \theta$. The latter is related to the
true temperature $\theta$ through the relation,
\begin{equation} \label{dilutethetabar}
\overline{\theta} \cong \theta \pm \frac{\rho}{16 \pi },
\end{equation}
where the $+$ and $-$ signs applies to fermions and bosons,
respectively.
\begin{figure}[htb]
\center
\includegraphics[width=0.8\linewidth]{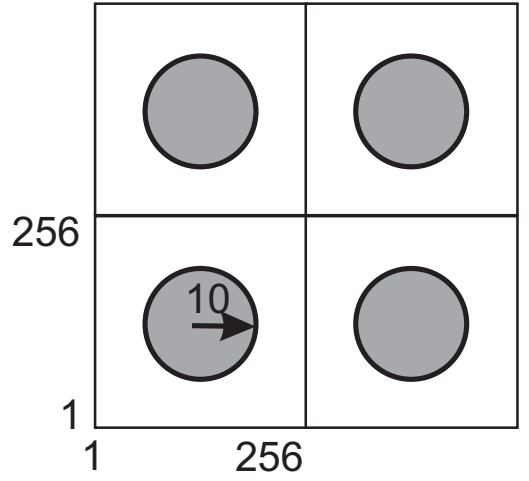}
\caption{Pictorial view of the initial condition taken in a
two-dimensional cell of 256$^2$ points with periodic boundary
conditions. The outward velocity (or the
density) takes distinct values inside and outside the circle of radius 10, while the remaining
variables are taken homogeneous throughout the cell.} \label{pbc}
\end{figure}
\begin{figure}[htb]
\center
\includegraphics[width=\linewidth]{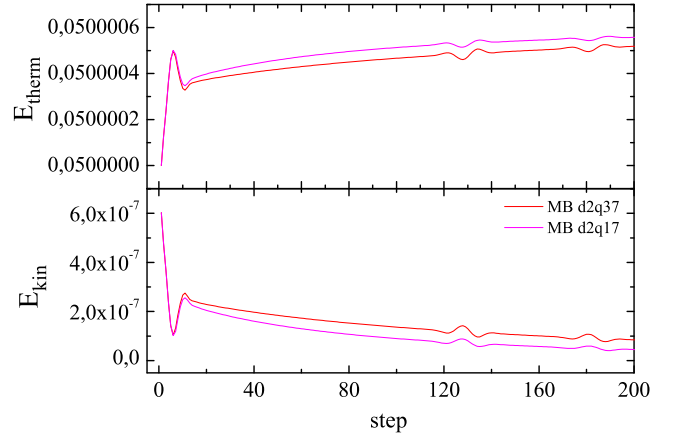}
\caption{The thermal and the kinetic energies for the Maxwell-Boltzmann distribution, with the circle of radial velocity as initial condition, are shown here for the first 200 time steps, using both the d2q17 and d2q37 lattices. For time step 118 the travelling front wave hits the side of the unit cell.} \label{vel-kintherm}
\end{figure}
\begin{figure}[htb]
\center
\includegraphics[width=\linewidth]{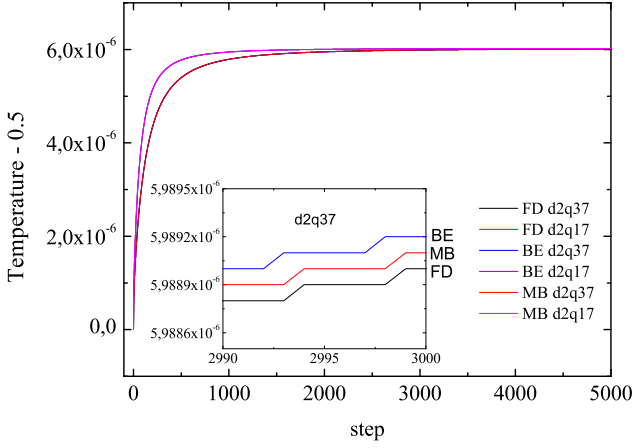}
\caption{The deviation of the temperature from its initial value is shown here for the first 5000 time steps using the circle of velocity as initial condition. The the splitting among the FD, MB and BE distribution curves is very small in comparison to  the d2q17 and d2q37 lattice splitting and for this reason not observable. The inset shows for a particular time step window the splitting of the temperature deviation among the three distributions for the d2q37 lattice.}
\label{vel-temp}
\end{figure}
\begin{figure}[htb]
\center
\includegraphics[width=\linewidth]{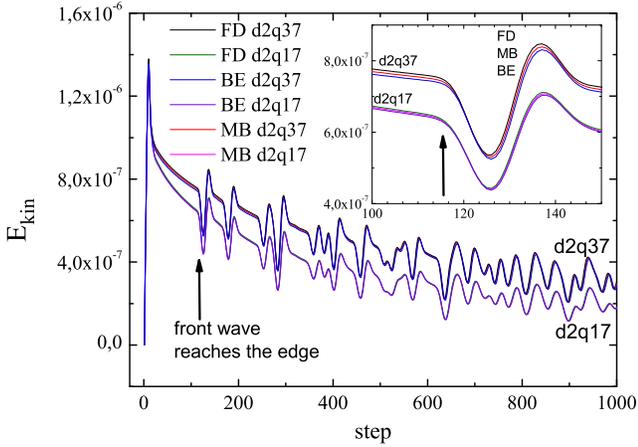}
\caption{The Kinetic energy of the FD, MB and BE distributions using the d2q17 and d2q37 lattices is shown here for the first 1000 time steps. The initial condition used is of the circle of density. The inset shows in a particular time step window the splitting of the kinetic energy among the three distributions for both d2q17 and d2q37 lattices.} \label{dens-kin}
\end{figure}
\begin{figure}[htb]
\center
\includegraphics[width=\linewidth]{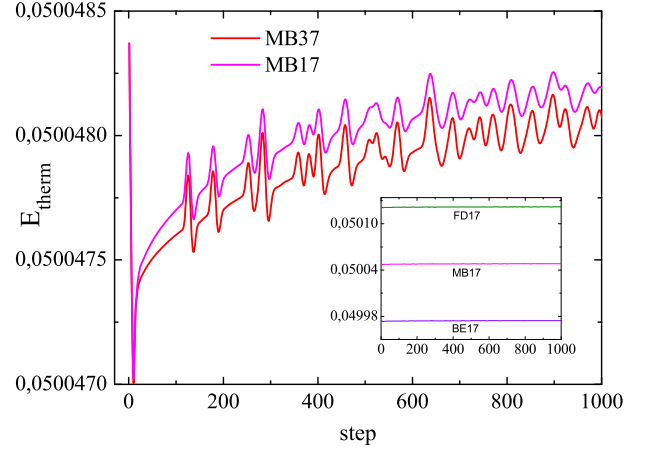}
\caption{The thermal energy of the MB  distribution using the d2q17 and d2q37 lattices is shown here for the first 1000 time steps. The initial condition used is that of the circle of density. The inset shows in a particular time step window the splitting of the kinetic energy among the three distributions for the d2q17 lattice.}
\label{dens-therm}
\end{figure}
\begin{figure}[htb]
\center
\includegraphics[width=\linewidth]{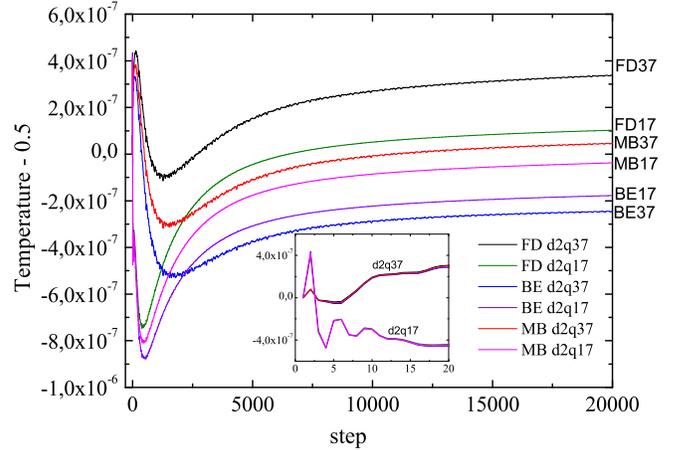}
\caption{The temperature deviation from its initial value is shown here for the case of a circle of density as initial condition for the first 20000 time steps. The curves for the three distributions and the two lattices are clearly seen here. The inset shows the first 20 time steps.} \label{dens-temp}
\end{figure}

\section{Numerical results}\label{numerical}
In this section we  use the LBM based on the present $N=4$ order
theory to numerically solve the dilute quantum fluid. We find that
the energy is indeed conserved in $D=2$ under periodic boundary
conditions and some initial conditions that trigger a time evolution
where motion and heat occur simultaneously. The initial conditions
are associated to a circle of radius $R$ located at the center of
the cell, where one of the variables, among the temperature
$\theta$, velocity $\boldsymbol{u}$, and the density $\rho$, assumes
a special value, whereas the other two are taken constant throughout
the cell. A pictorial view of such initial conditions is shown in
Fig.~\ref{pbc}. We work with a grid of $A_{cell}=256^2$ points, and
in these units $R=10$. We take for the time step of
Eq.(\ref{disboltzeq0}) $\tau/\Delta t=0.58$. We use in this section the concept of a real temperature $T$ connected to the reduced one by
$T \equiv c_s^2\theta$, where $c_s$ is a distinct numerical parameter
for each of the lattices d2q17 and d2q37, This parameter makes equal to one
the smallest non-zero microscopic velocity along the
axis~\cite{PhysRevE.77.026707}. In dimensionless units $c_s$ plays
the role of the speed $c_r$, defined by Eq.(\ref{reduced-velocity}).
The time evolution of the system is a multiple of the computer steps performed in our numerical procedure.
Since our goal is to verify that the total
energy remains constant in the cell at any time step, we define the
following sums over all cell points, which are the average values of
the kinetic and the thermal energies in the cell.
\begin{eqnarray}
&& E_{kin}\equiv \frac{1}{A_{cell}}\sum_{cell} \frac{1}{2}\rho \boldsymbol{u}^2 \label{ekin}\\
&& E_{therm}\equiv \frac{1}{A_{cell}}\sum_{cell} \frac{D}{2}\rho \bar \theta, \, \mbox{and} \label{etherm}\\
&& E_{total} \equiv  E_{kin}+E_{therm},\label{etotal}
\end{eqnarray}
In our numerical study we assume identical initial conditions for
the BE, FD and MB cases, which means that they start with the same
density, temperature and velocity. Thus the initial  kinetic energy
of Eq.(\ref{ekin}) is the same for the three cases, but not the
thermal energy of Eq.(\ref{etherm}) since the initial pseudo
temperature is not the same for the three cases, according to
Eq.(\ref{dilutethetabar}), being  greater for fermions and smaller
for bosons, as compared to the classical case.

To perform numerical calculations we must employ a concrete
realization of the weights and microscopic velocities that satisfy
the algebra of Eqs.(\ref{sum0}) to (\ref{sum6}). This assures that
the Gauss-Hermite quadrature is being correctly performed. We choose
the so-called d2q17~\cite{FLM:409537} and
d2q37~\cite{PhysRevE.77.026707} microscopic velocity lattices. It is
important to make a few considerations about these lattices
regarding their ability to give the expected answers. This ability
relies on the order of the maximum power of the microscopic velocity
polynomial contained in the equilibrium distribution function and
also the calculated moments. The equilibrium distribution function
has maximum order ${f}^{(0)}_{\alpha}\sim \xi_{\alpha}^N$, for $N=3$
and $N=4$, as seen from Eq.(\ref{edf-cont3}) and Eq.(\ref{edf-dis}),
respectively. For computer purposes the highest moment that must be
calculated is the energy, which is of order $\xi_{\alpha}^2$. Thus
the chosen microscopic velocity lattice must correctly account for
the sums of the kind Eqs.(\ref{sum0}) to (\ref{sum6}), which means
$\xi_{\alpha}^{5}$ for $N=3$ and $\xi_{\alpha}^{6}$ for $N=4$.
According to Ref.~\onlinecite{FLM:409537} these calculations can be
correctly done in the d2q9 and d2q17 algebras for $N=3$ and $N=4$,
respectively. Thus we conclude that d2q17 and d2q37 can be used for
the numerical treatment of the present $N=4$ order LBM. Nevertheless
to apply the Chapman-Enskog analysis and derive the equation for
energy, one must compute the moment $\phi^{jk}\sim
\xi_{\alpha}^{4}{f}^{(0)}_{\alpha}$. Hence in this case the
microscopic velocity lattice must correctly account for sums beyond
Eqs.(\ref{sum0}) to (\ref{sum6}), which means of the order
$\xi_{\alpha}^{7}$ for $N=3$ and $\xi_{\alpha}^{8}$ for $N=4$. Thus
for this purpose the d2q17 lattice is limited to
$N=3$~\cite{FLM:409537} and cannot be used $N=4$. For such purpose
$N=4$ demands a higher order lattice, such as
d2q37~\cite{PhysRevE.77.026707}.

\subsection{Velocity initial conditions}\label{vel}
Under this initial condition $\boldsymbol{u}$ is non-zero inside a
circle of radius $R$, where the outward component is equal to 0.02
in reduced units, being zero outside the circle. Thus at the
beginning radial motion is set for a dilute quantum fluid of
constant density, $\rho=0.1$, and temperature, $T=0.5$, in the unit
cell.
This initial motion immediately generates friction and  subsequent
heating that raises the temperature. Fig.~\ref{vel-kintherm} shows
the time step evolution  of $E_{kin}$ and $E_{therm}$ and the most
important aspect found in these two plots is that their sum is a
constant which is $E_{total}$, as given by Eq.(\ref{etotal}).
$E_{kin}$ follows the trend of decrease in time, and eventually must
vanish, while $E_{therm}$ increases and tends to stabilize, though
such behaviours are just suggested in the first 200 steps of
Fig.~\ref{vel-kintherm}. This figure only intends to display the
evolution immediately  after the initial condition. To reach a
nearly steady state of zero velocity and stable temperature
simulations up to 20.000 steps must be carried. The two employed
lattices, namely d2q17 and d2q37, give the same qualitative results
for the time evolution of the system, but their numerical difference
hinders the splitting among the BE, FD and MB curves for each
lattice as they fall very close to each other. Notice the non-zero
initial value of $E_{kin}$, due to the initial velocity condition.
After a few steps $E_{kin}$ drops to a local minimum, which is a
local maximum for $E_{therm}$, meaning that friction caused by
motion raises the temperature but in such a way that $E_{total}$ is
conserved. After 118 steps the front wave raised by the initial
condition reaches the side of the unit cell. Notice that this number
coincides with the number of grid points between the circle and the
unit cell side, namely, (256-20)/2. In the numerical algorithm the
speed of propagation is one,  thus the number of steps required for
the signal reach the border is the number of grid points itself.
Indeed Fig.~\ref{vel-kintherm} shows a hilly behaviour starting
nearly at 120 steps, a consequence of the interference of incoming
waves from the neighbour cells. Fig.~\ref{vel-temp} shows the
evolution of the temperature deviation from its initial value of
$T=0.5$. After 5000 steps the d2q17 and the d2q37 converge basically
to the same final temperature which is just $6.0\; 10^{-6}$ above
the initial one. Interestingly there is also oscillatory behaviour
in the deviated temperature though much less perceptible than in
Fig.~\ref{vel-kintherm}. The inset of Fig.~\ref{vel-temp} shows the
splitting between the BE, FD and MB cases in a particular time step
window. Indeed one observes that this splitting between the three
statistics is of order $10^{-9}$ whereas the differences between
d2q17 and d2q37 is of order $10^{-6}$.

\subsection{Density initial conditions}\label{dens}
Under this initial condition the density is $\rho=0.12$ inside and
$\rho=0.1$ outside the circle of radius $R$. The initial temperature
is $T=0.5$ and the macroscopic velocity  $\boldsymbol{u}$ is zero
everywhere. Though there is no initial motion, the non homogeneous
density distribution sets a pressure front. The higher density at
the center means higher pressure that forces motion outward the
circle, and so, generates friction, heating and raise of the
temperature in the cell. The first 1000 steps of $E_{kin}$ are shown
in Fig.~\ref{vel-kintherm}. Notice that it starts from zero and the
oscillatory behavior set by the entrance of wave fronts from
neighbor cells is well described by the d2q17 and d2q37 lattices
since their numerical  differences are in the range of $10^{-7}$. In
this case too the difference among the three distributions is hidden
by the lattice difference, shown in the inset for a chosen window of
time steps for both lattices. The trend towards kinetic energy decay
is seen in this figure though the convergence towards a steady state
is only suggested by  this figure. To really see it a larger window
must be taken. The time step evolution of $E_{therm}$ is shown in
Fig.(\ref{dens-therm}). $E_{therm}$ experiments a sudden drop to
allow for the increase of $E_{kin}$ as the system starts motion due
to pressure unbalance. As in the previously case, $E_{total}$ is
absolutely conserved to machine precision, in our case tested to
order $10^{-10}$. The inset of this plot shows the splitting of the
three distributions within the first 1000 steps. Notice that such
values are distinct because we chose the same initial $\theta$ for
the three distributions, which means different $\bar \theta$ in the
three cases. Finally we address the evolution of the temperature
deviation from its initial value $T=0.5$ in Fig.(\ref{dens-temp})
within the full time step window studied, namely 20000 steps. The
very first 20 steps, shown in the inset, indicate that the initial
temperature is the same in all cases. However the final ones are
different. The d2q17 reaches a final temperature deviation  at a
value slightly lower that the d2q37 case by an order of magnitude of
$10^{-7}$. The final deviation is higher for the FD distribution and
lower for the BE. We notice during evolution a drop to a minimum of
temperature required to accommodate  the kinetic expansion caused by
the pressure unbalance in an energy conserving scenario.

\section{Conclusion}\label{conclusion}
We have conclusively shown that the Hermite polynomial expansion of
the equilibrium distribution functions must be carried to fourth
order for quantum fluids. Only in this order it is possible to
obtain meaningful macroscopic hydrodynamical equations that lead to
the correct viscosity and thermal coefficients. We have also
demonstrated the feasibility of the fourth order lattice Boltzmann
method scheme by showing that it numerically describes motion and
heating in an energy conserving way.
\begin{acknowledgements}
Anderson Ilha,  R. M. Pereira, and Valter Yoshihiko Aibe
thank to Inmetro for financial support.  M. M. Doria acknowledges the Brazilian agency CNPq and Inmetro for financial support. Rodrigo C. V. Coelho thanks to CNPq.
\end{acknowledgements}
\appendix
\section{Hermite polynomials}\label{appendix-hermite}
The lowest order Hermite polynomials are easily derived from
Eq.(\ref{Hpoly}):
\begin{eqnarray}
&&H(\boldsymbol{\xi}) = 1 \label{H0} \\
&&H^i(\boldsymbol{\xi})=\xi^i\label{H1}  \\
&&H^{ij}(\boldsymbol{\xi})=\xi^i \xi^j - \delta^{ij} \label{H2} \\
&&H^{ijk}(\boldsymbol{\xi})=\xi^i \xi^j \xi^k - (\xi^i \delta^{jk} +
\xi^j \delta^{ik} + \xi^k \delta^{ij}) \label{H3} \\
&&H^{ijkl}(\boldsymbol{\xi})=\xi^i \xi^j \xi^k \xi^l - (\xi^i\xi^j
\delta^{kl} + \xi^i \xi^k \delta^{jl} + \xi^i\xi^l \delta^{jk} +
\nonumber \\
&&+ \xi^j \xi^k \delta^{il} + \xi^j\xi^l \delta^{ik} + \xi^k\xi^l
\delta^{ij})+\delta^{ijkl}.\label{H4}
\end{eqnarray}
The following integrals over the gaussian function of
Eq.(\ref{omega0}) are a consequence of the orthonormality of the Hermite polynomials:
\begin{eqnarray}
&& \int d^D \boldsymbol{\xi}\, \omega(\boldsymbol{\xi})=1
\label{int0} \\
&& \int d^D \boldsymbol{\xi}\,
\omega(\boldsymbol{\xi})\xi^i =0 \label{int1} \\
&& \int d^D \boldsymbol{\xi}\,
\omega(\boldsymbol{\xi})\xi^i \xi^j= \delta^{i j} \label{int2} \\
&& \int d^D \boldsymbol{\xi}\,
\omega(\boldsymbol{\xi})\xi^i \xi^j \xi^k= 0 \label{int3} \\
&& \int d^D \boldsymbol{\xi}\, \omega(\boldsymbol{\xi})\xi^i\xi^j\xi^k\xi^l
= \delta^{ijkl}\label{int4} \\
&& \int d^D \boldsymbol{\xi}\,
\omega(\boldsymbol{\xi})\xi^i\xi^j\xi^k\xi^l\xi^m= 0 \label{int5} \\
&& \int d^D \boldsymbol{\xi}\,
\omega(\boldsymbol{\xi})\xi^i\xi^j\xi^k\xi^l\xi^m \xi^n= \delta^{ijklmn}.
\label{int6}
\end{eqnarray}
The $\delta^{\cdots}$ tensors are constructed recursively from the Kronecker´s symbol:
$\delta^{ij}=1$ for $i=j$ and 0 for $i\neq j$. In case of four
indices,
\begin{eqnarray} \label{delta-4}
\delta^{ijkl} \equiv \delta^{ij}\delta^{kl}+\delta^{ik}\delta^{jl} + \delta^{il} \delta^{kj}.
\end{eqnarray}
The sixth order tensor is,
\begin{eqnarray}\label{delta-6}
&& \delta^{i j k l m n } \equiv \delta^{i j}\delta^{k l m n}+
\delta^{i k}\delta^{j l m n}+ \nonumber \\
&&+\delta^{i l}\delta^{k j m n}+ \delta^{i
m}\delta^{j k l n}+ \delta^{i n}\delta^{j k l m}.
\end{eqnarray}
The eight order is,
\begin{eqnarray}\label{delta-8}
&&\delta^{ijklmnpq} \equiv \delta^{ij}\delta^{klmnpq} +\delta^{ik} \delta^{jlmnpq}+ \delta^{il}\delta^{jkmnpq} +
\nonumber \\
&&+\delta^{im} \delta^{jklnpq}+ \delta^{in}\delta^{jklmpq}+ \delta^{ip} \delta^{jklmnq}+ \delta^{iq} \delta^{jklmnp},\nonumber \\
\end{eqnarray}
and so forth.

To illustrate that Eqs.(\ref{int1}-\ref{int6}) are a consequence of
Eq.(\ref{hermiteortho}) let us consider two examples. The
orthogonality of the first two Hermite polynomials is,
\begin{equation}\label{ortho-01}
\int d^D \boldsymbol{\xi}\, \omega(\xi) H(\boldsymbol{\xi})
H^i(\boldsymbol{\xi}) = \int d^D \boldsymbol{\xi}\, \omega(\xi) \xi^i = 0,
\end{equation}
which yields Eq.(\ref{int1}). Integration over two first order
Hermite polynomials gives,
\begin{equation}
\int d^D \boldsymbol{\xi}\, \omega(\xi) H^i(\boldsymbol{\xi})
H^j(\boldsymbol{\xi}) = \int d^D \boldsymbol{\xi}\, \omega(\xi) \xi^i \xi^j
= \delta^{ij},
\end{equation}
which is Eq.(\ref{int2}). Integration over the first and the second
order Hermite polynomials gives that,
\begin{eqnarray}
\int d^D \boldsymbol{\xi}\, \omega(\xi) H^i(\boldsymbol{\xi})
H^{jk}(\boldsymbol{\xi}) = \\
= \int d^D \boldsymbol{\xi}\, \omega(\xi) \xi^i \left(\xi^j
\xi^k-\delta^{jk} \right) =0.
\end{eqnarray}
From Eq.(\ref{ortho-01}) one obtains Eq.(\ref{int3}). As a last
example we consider the integration over two second order Hermite
polynomials:
\begin{eqnarray} &&\int d^D \boldsymbol{\xi}\,
\omega(\boldsymbol{\xi}) H^{ij}(\boldsymbol{\xi})H^{kl}(\boldsymbol{\xi})
= \nonumber \\
&& = \int d^D \boldsymbol{\xi}\, \omega(\boldsymbol{\xi})[\xi^i \xi^j - \delta^{ij}][\xi^k \xi^l - \delta^{kl}] = \nonumber \\
&& = \int d^D \boldsymbol{\xi}\, \omega(\boldsymbol{\xi}) \xi^i \xi^j \xi^k \xi^l - \delta^{ij} \int d^D \boldsymbol{\xi}\, \omega(\boldsymbol{\xi}) \xi^k \xi^l - \nonumber \\
&& - \delta^{kl} \int d^D \boldsymbol{\xi}\, \omega(\boldsymbol{\xi}) \xi^i \xi^j + \delta^{ij} \delta^{kl} \int d^D \boldsymbol{\xi}\, \omega(\boldsymbol{\xi})= \nonumber \\
&& = \delta^{ik} \delta^{jl}+ \delta^{il}\delta^{kj}. \nonumber
\end{eqnarray}

The orthogonality relations, given by Eqs.(\ref{int1})-(\ref{int6})
are the key ingredients to obtain the moments of Eqs.(\ref{fbar1}),
(\ref{fbar2}) and (\ref{fbar3}).

It is worth to investigate the properties of  the Hermite polynomials in case the argument is  a difference $\xi-u$. Using (\ref{H0}), (\ref{H1}),
(\ref{H2}), (\ref{H3}) we find the following relations:
\begin{eqnarray}
&&H^{i}(\boldsymbol{\xi})=H^{i}(\boldsymbol{\xi-u})+ u^i H
\label{hermite-with-u-1}
\end{eqnarray}
\begin{eqnarray}
&&H^{ij}(\boldsymbol{\xi})= H^{ij}(\boldsymbol{\xi-u}) +
u^iH^j(\boldsymbol{\xi-u})+ \nonumber \\
&&+u^j H^i(\boldsymbol{\xi-u})+ u^iu^jH \label{hermite-with-u-2}
\end{eqnarray}
\begin{eqnarray}
&&H^{ijk}(\boldsymbol{\xi})=H^{ijk}(\boldsymbol{\xi-u})+ u^i
H^{jk}(\boldsymbol{\xi-u})+ \nonumber \\
&&+u^jH^{ik}(\boldsymbol{\xi-u}) + u^k
H^{ij}(\boldsymbol{\xi-u})+u^iu^jH^k(\boldsymbol{\xi-u})+ \nonumber
\\
&&+u^iu^kH^j(\boldsymbol{\xi-u})+u^ju^kH^i(\boldsymbol{\xi-u})+ u^iu^ju^k H
\label{hermite-with-u-3}
\end{eqnarray}
\begin{eqnarray}
&&H^{ijkl}(\boldsymbol{\xi})=H^{ijkl}(\boldsymbol{\xi-u})+ u^i
H^{jkl}(\boldsymbol{\xi-u})+ \nonumber \\
&&+ u^j H^{ikl}(\boldsymbol{\xi-u})+u^k H^{ijl}(\boldsymbol{\xi-u})+u^l H^{ijk}(\boldsymbol{\xi-u})+\nonumber \\
&&+u^iu^jH^{kl}(\boldsymbol{\xi-u})+u^iu^kH^{jl}(\boldsymbol{\xi-u}) +u^iu^lH^{jk}(\boldsymbol{\xi-u})+\nonumber\\
&&+u^ku^lH^{ij}(\boldsymbol{\xi-u})+u^ju^lH^{ik}(\boldsymbol{\xi-u}) +u^ju^kH^{il}(\boldsymbol{\xi-u})+\nonumber\\
&&+u^iu^ju^kH^{l}(\boldsymbol{\xi-u})+u^iu^ju^lH^{k}(\boldsymbol{\xi-u}) +\nonumber\\
&&+u^ju^ku^lH^{i}(\boldsymbol{\xi-u})+u^iu^ku^lH^{j}(\boldsymbol{\xi-u}) +\nonumber\\
&&+u^iu^ju^ku^l. \label{hermite-with-u-4}
\end{eqnarray}
Using the above expressions it becomes straightforward to calculate
the first four coefficients of the Hermite expansion.

\section{Taylor expansion of the Maxwell-Boltzmann distribution function}\label{appendix-Taylor}
In this Appendix we consider the Maxwell-Boltzmann equilibrium distribution in reduced units,
\begin{eqnarray}\label{fprime}
f(\boldsymbol{\xi}) \equiv \rho\left( \frac{1}{2 \pi \theta }
\right)^{\frac{D}{2}} e^{-\frac{(\boldsymbol{\xi}-\boldsymbol{u})^2}{2
\theta}},
\end{eqnarray}
whose first three moments are obtained as below.
\begin{eqnarray}
\rho(\boldsymbol{x})&=& \int d^D \boldsymbol{\xi} f(\boldsymbol{\xi}), \label{redmoment1}\\
\boldsymbol{u}(\boldsymbol{x}) &=&\frac{1}{\rho(\boldsymbol{x})}\int
d^D \boldsymbol{\xi} \; \boldsymbol{\xi} f(\boldsymbol{\xi}), \, \mbox{and}, \label{redmoment2}\\
\frac{D}{2} \theta(\boldsymbol{x}) &=&\frac{1}{\rho(\boldsymbol{x})}\int d^D
\boldsymbol{\xi}\;\frac{1}{2}
\left[\boldsymbol{\xi}-\boldsymbol{u}(\boldsymbol{x})\right]^2
f(\boldsymbol{\xi}).\label{redmoment3}
\end{eqnarray}
We seek a Taylor expansion of the Maxwell-Boltzmann equilibrium
distribution function such that the three above moment relations are
retained for the expanded distribution function. The reference
temperature and velocity are the parameters that set the Taylor
expansion, although they are not present in the original
distribution. This means that the limit that the macroscopic
velocity and the temperature deviation are small in comparison to
$c_r$ and $T_r$, respectively, is being considered. Therefore
$\boldsymbol{u}(\boldsymbol{x})$ and $\theta(\boldsymbol{x})-1$ are
small quantities, fact that justifies a series expansion in powers
of these quantities. The present Taylor expansion  makes evident the
presence of a small parameter $\epsilon$ and we take the expansion
until order $\epsilon^2$. The square of the macroscopic velocity and
the temperature deviation must be considered of the same order:
$\vert \boldsymbol{u}(\boldsymbol{x})\vert \sim \sqrt{\epsilon}$ and
$\theta(\boldsymbol{x})-1 \sim \epsilon$. A few remarks are worth of
notice. The only scalars available are $\boldsymbol{\xi \cdot u}$
and $\boldsymbol{u}^2$. The microscopic velocity $\boldsymbol{\xi}$
has no $\epsilon$ order assigned to it being limited to small values
by the gaussian decay. Therefore according to our expansion
criterion the only terms to be kept are those proportional to 1,
$\boldsymbol{\xi \cdot u}$, $\boldsymbol{u}^2$, $(\boldsymbol{\xi
\cdot u})^2$, $\boldsymbol{u}^3$, $(\boldsymbol{\xi \cdot
u})^2\boldsymbol{u}$, $(\boldsymbol{\xi \cdot u})\boldsymbol{u}^2$,
$(\boldsymbol{\xi \cdot u})^3$, $\boldsymbol{u}^4$,
$(\boldsymbol{\xi \cdot u})^2\boldsymbol{u}^2$, $(\boldsymbol{\xi
\cdot u})^4$, $(\theta-1)$, $(\theta-1)\boldsymbol{\xi \cdot u}$,
$(\theta-1)\boldsymbol{u}^2$, $(\theta-1)(\boldsymbol{\xi \cdot
u})^2$, and $(\theta-1)^2$. Firstly consider small deviations of
$T_r$, namely, $\theta = 1+ (\theta -1)$ up to order $\epsilon^2$ in
the denominator of Eq.(\ref{fprime}):
\begin{eqnarray}
&&\frac{1}{\theta^{\frac{D}{2}}} =
\frac{1}{[1+(\theta-1)]^{\frac{D}{2}}} = \nonumber
\\
&&=1-\frac{D}{2}(\theta -1) +\frac{D}{4}\frac{D+2}{2} (\theta-1)^2
+O(\epsilon^3). \label{expansion_theta}
\end{eqnarray}
The exponential also has a $\theta$ dependent denominator that must
be expanded resulting in three different terms, which must be treated
separately according to $\epsilon$.
\begin{eqnarray}\label{exp3}
& & \exp \left (- \frac{(\boldsymbol{\xi}-\boldsymbol{u})^2}{2 \theta} \right ) = \nonumber \\
&=& \exp \left\{ -\frac{(\boldsymbol{\xi}-\boldsymbol{u})^2}{2}\left[ 1-(\theta-1)+(\theta-1)^2 +O(\epsilon^3) \right] \right\} = \nonumber \\
&=& \exp\left[-\left( \frac{\boldsymbol{\xi}^2}{2} -\boldsymbol{\xi \cdot u}+ \frac{\boldsymbol{u}^2}{2} \right) \right] \cdot \nonumber \\
& & \cdot \exp \left[ \left( \frac{\boldsymbol{\xi}^2}{2} -\boldsymbol{\xi \cdot u}+ \frac{\boldsymbol{u}^2}{2} \right)(\theta-1) \right] \cdot \nonumber \\
& & \cdot \exp \left[- \left( \frac{\boldsymbol{\xi}^2}{2} -\boldsymbol{\xi
\cdot u}+ \frac{\boldsymbol{u}^2}{2} \right)(\theta-1)^2 \right] \ldots
\end{eqnarray}
In the first exponential we factorize the gaussian function (Eq.(\ref{omega0})), $\exp\left(
-\boldsymbol{\xi}^2/2 \right)$, because of its zeroth  $\epsilon$ order.
Next we select for the three exponentials only those terms of order
equal or lower than $\epsilon^2$. The first exponential becomes
\begin{eqnarray}
&&\exp\left[-\left( \frac{\boldsymbol{\xi}^2}{2} -\boldsymbol{\xi
\cdot u}+ \frac{\boldsymbol{u}^2}{2} \right) \right] = \exp{\left(-\frac{\boldsymbol{\xi}^2}{2} \right)} \cdot \nonumber \\
&& [ 1+ \boldsymbol{\xi \cdot u} - \frac{\boldsymbol{u}^2}{2}+
\frac{1}{2}(\boldsymbol{\xi \cdot u})^2 -
\frac{1}{2}(\boldsymbol{\xi \cdot u}) \boldsymbol{u}^2 + \label{first_exponential} \\
&& \frac{\boldsymbol{u}^4}{8} + \frac{1}{6} (\boldsymbol{\xi \cdot u})^3 -
\frac{1}{4}(\boldsymbol{\xi \cdot u})^2 \boldsymbol{u}^2 +
\frac{1}{24}(\boldsymbol{\xi \cdot u})^4 +O(\epsilon^{5/2})].\nonumber
\end{eqnarray}
The second one,
\begin{eqnarray}
&&\exp\left[-\left( \frac{\boldsymbol{\xi}^2}{2} -\boldsymbol{\xi \cdot u}+
\frac{\boldsymbol{u}^2}{2} \right)(\theta -1) \right] =1+
\label{second_exponential} \\
&& + \left( \frac{\boldsymbol{\xi}^2}{2} -\boldsymbol{\xi \cdot u}+
\frac{\boldsymbol{u}^2}{2} \right)(\theta-1)+ \frac{\boldsymbol{\xi}^4}{8}
(\theta-1)^2+O(\epsilon^{5/2}),\nonumber
\end{eqnarray}
and the third,
\begin{eqnarray}
&& \exp \left[ - \left( \frac{\boldsymbol{\xi}^2}{2} -\boldsymbol{\xi \cdot
u}+ \frac{\boldsymbol{u}^2}{2}
\right)(\theta-1)^2 \right] = \label{third_exponential} \\
&=&
1-\frac{\boldsymbol{\xi}^2}{2}(\theta-1)^2+O(\epsilon^{5/2}).\nonumber
\end{eqnarray}
The product of Eqs.(\ref{expansion_theta}) and (\ref{exp3}),
together with the expansions of Eq.(\ref{first_exponential}),
(\ref{second_exponential}), (\ref{third_exponential}), gives that
\begin{eqnarray}\label{func0}
&& f=\rho \omega(\boldsymbol{\xi}) \left[ 1-\frac{D}{2} (\theta-1) +
\frac{D}{4}\frac{D+2}{2}(\theta-1)^2 -O(\epsilon^3)\right]\cdot \nonumber \\
&& \cdot \big[1+\boldsymbol{\xi \cdot u} -
\frac{\boldsymbol{u}^2}{2}+\frac{1}{2}(\boldsymbol{\xi \cdot u})^2 -
\frac{1}{2}(\boldsymbol{\xi \cdot u})\boldsymbol{u}^2+
\frac{\boldsymbol{u}^4}{8} + \nonumber \\
&& +\frac{1}{6}(\boldsymbol{\xi \cdot \vec u})^3  -
\frac{1}{4}(\boldsymbol{\xi \cdot u})^2 \boldsymbol{u}^2 + \frac{1}{24}
(\boldsymbol{\xi \cdot u})^4-O(\epsilon^{5/2}) \big] \cdot \big[ 1+ \nonumber \\
&& +\left(\frac{\boldsymbol{\xi}^2}{2}- \boldsymbol{\xi \cdot u}  +
\frac{\boldsymbol{u}^2}{2} \right)(\theta-1) +
\frac{\boldsymbol{\xi}^4}{8}(\theta-1)^2 +O(\epsilon^{5/2}) \big]\cdot \nonumber \\
&& \cdot \big[1- \frac{\boldsymbol{\xi}^2}{2}(\theta-1)^2
+O(\epsilon^{5/2}) \big]
\end{eqnarray}
where we have used the gaussian function of Eq.(\ref{omega0}).
Finally we  expand Eq.(\ref{func0}) and hold terms up to  order
$\epsilon^2$, by multiplying the $\epsilon$ expansions of the three
exponentials of the numerator with that of the denominator:
\begin{eqnarray} \label{fTaylor}
&&\frac{\bar f}{\rho \omega(\boldsymbol{\xi})}= 1+ \boldsymbol{\xi \cdot u}
+ \frac{1}{2}(\boldsymbol{ \xi \cdot u} )^2 - \frac{u^2}{2}
+\frac{1}{2}(\theta-1)(\boldsymbol{\xi}^2 -D)+ \nonumber \\
&&+\frac{1}{6}(\boldsymbol{\xi \cdot u})^3 -
\frac{1}{2}\boldsymbol{u}^2(\boldsymbol{\xi \cdot u}) +
\frac{1}{2}(\theta-1)(\boldsymbol{\xi \cdot u})(\boldsymbol{\xi}^2
-D-2) + \nonumber \\
&&+\frac{1}{24}(\boldsymbol{\xi \cdot u})^4 -\frac{1}{4} (\boldsymbol{\xi
\cdot u})^2 \boldsymbol{u}^2 + \frac{1}{8}\boldsymbol{u}^4+
\nonumber \\
&& + \frac{1}{4}(\theta -1)\big[ (\boldsymbol{\xi \cdot u})^2
(\boldsymbol{\xi}^2 -D-4) + \boldsymbol{u}^2(D+2-\boldsymbol{\xi}^2)
\big] + \nonumber \\
&&+ \frac{1}{8}(\theta-1 )^2 \big[
\boldsymbol{\xi}^4-2(D+2)\boldsymbol{\xi}^2 + D(D+2) \big]+O(\epsilon^{5/2})
\end{eqnarray}
We have obtained a Taylor expansion of the Maxwell Boltzmann equilibrium distribution function
(Eq.(\ref{fprime})) in powers of $T(\boldsymbol{x})/T_r-1$ and
$\boldsymbol{v(\boldsymbol{x})}/c_r$ up to the desired order of
$\epsilon^2$. Remarkably,  the three local parameters $\rho$,
$\theta$, and $\boldsymbol{u}$, contained in $\bar f$, are also its
first three moments. The above Taylor expansion
also satisfies the same relations for the momentum flux tensor,
$\int d^D \boldsymbol{\xi}\, \xi^i \xi^j\, \bar
f(\boldsymbol{\xi})=\delta^{ij}\rho\theta D+\rho u^i u^j$, and for the
energy flux tensor, $\int d^D \boldsymbol{\xi}\, \xi^i \boldsymbol{\xi}^2\,
\bar f(\boldsymbol{\xi})=[\delta^{ij}\rho\theta(D+2)/2+\rho
\boldsymbol{u}^2/2]u^i$.
\begin{equation}
\rho(\boldsymbol{x})= \int d^D \boldsymbol{\xi} \bar
f(\boldsymbol{\xi},\rho,\theta,\boldsymbol{u}), \label{fbar1}
\end{equation}
\begin{equation}
\boldsymbol{u}(\boldsymbol{x}) =\frac{1}{\rho(\boldsymbol{x})}\int d^D
\boldsymbol{\xi} \; \boldsymbol{\xi} \bar
f(\boldsymbol{\xi},\rho,\theta,\boldsymbol{u}), \label{fbar2}
\end{equation}
and,
\begin{equation}
\frac{D}{2} \theta(\boldsymbol{x}) =\frac{1}{\rho(\boldsymbol{x})}\int d^D
\boldsymbol{\xi}\;\frac{\left[\boldsymbol{\xi}-\boldsymbol{u}(\boldsymbol{x})\right]^2}{2}
\bar f(\boldsymbol{\xi},\rho,\theta,\boldsymbol{u}). \label{fbar3}
\end{equation}

\section{The position and time cross derivative terms of the momentum and energy conservation equations }\label{appendix-cross-derivative}

In this Appendix we seek replacement of the
$\frac{\partial}{\partial x^j}\frac{\partial}{\partial t}$ terms
contained in Eqs. (\ref{eq_momentum_with_crossing_derivative}) and
(\ref{macro-e2}) by position derivatives. The key to this step is to
consider that terms proportional to the square of the collision time
$\tau$, namely, proportional to $[\tau ( 1-\Delta t/2 \tau)]^2$, are
neglected. Thus it consider the above equations in the limit $\tau
\rightarrow 0$ and $\Delta t/\tau$ fixed. Setting $\tau =0$ in Eqs.
(\ref{eq_momentum_with_crossing_derivative}) and (\ref{macro-e2})
gives that,
\begin{eqnarray}
&& \frac{\partial}{\partial t }\left(\rho u^i\right) + \frac{\partial}{\partial x^j} \left( \rho \bar \theta \delta^{ij} + \rho u^i u^j  \right) =0, \label{eq_momentum_0th_order}\\
&&\frac{\partial}{\partial t} \left( \frac{\rho}{2}\boldsymbol{u}^2 + \frac{\rho}{2} D\bar \theta  \right) + \frac{\partial}{\partial x^j} \left[ \left( \frac{\rho}{2}\boldsymbol{u}^2 + \frac{\rho}{2}\bar \theta  (D+2)\right) u^j\right] =0. \nonumber \\ \label{eq_energy_0th_order}
\end{eqnarray}
\subsection{The momentum conservation equation}
Introducing the continuity equation (Eq.(\ref{continuity}))
into Eq. (\ref{eq_momentum_0th_order}) gives,
\begin{equation}
\rho \frac{\partial u^i}{\partial t} + \frac{\partial}{\partial x^j} \left( \rho \bar \theta   \delta^{ij} \right) + \rho u^j \frac{\partial u^i}{\partial x^j} =0.
\label{eq_momentum+continuity}
\end{equation}
We develop some identities from Eqs.(\ref{continuity}) and,
(\ref{eq_momentum+continuity}).
\begin{eqnarray}
&&\frac{\partial}{\partial t}(\rho \boldsymbol{u}^2) = \boldsymbol{u}^2 \frac{\partial \rho}{\partial t} + 2\rho u^j\frac{\partial u^j}{\partial t} = \nonumber \\
&&= -\boldsymbol{u}^2 \frac{\partial}{\partial x^j} (\rho u^j) -2u^j \frac{\partial}{\partial x^i} \left( \rho\bar \theta \delta^{ij} \right) - u^j \frac{\partial}{\partial x^j}(\rho\boldsymbol{u}^2) =\nonumber \\
&&= -\frac{\partial}{\partial x^j} (\rho\boldsymbol{u}^2 u^j) - 2u^j \frac{\partial}{\partial x^j} \left( \rho\bar \theta \right). \label{d/dtnu^2}
\end{eqnarray}
Introducing  Eq.(\ref{d/dtnu^2}) into Eq.(\ref{eq_energy_0th_order}) gives the identity,
\begin{equation*}
\frac{\partial}{\partial t} \left(\rho\bar \theta D \right) - 2u^j \frac{\partial}{\partial x^j} \left( \rho\bar \theta \right) + \frac{\partial}{\partial x^j} \left[ \rho\bar \theta (D+2) u^j \right] =0,
\end{equation*}
that can be expressed as,
\begin{equation}
\frac{\partial}{\partial t}\left( \rho\bar \theta D \right) =- \frac{\partial}{\partial x^j} \left( \rho\bar \theta D  u^j \right) -2\rho\bar \theta \frac{\partial u^j}{\partial x^j}.
\label{d/dtnthetaDg/g}
\end{equation}
Starting from,
\begin{equation}
\frac{\partial}{\partial t}(\rho u^i u^j) = u^i u^j \frac{\partial \rho}{\partial t}+ \rho u^j \frac{\partial u^i}{\partial t} + \rho u^i \frac{\partial u^j}{\partial t},
\end{equation}
and using the continuity equation, Eq.(\ref{continuity}), and
Eq.(\ref{eq_momentum+continuity}), yields that,
\begin{eqnarray}
&&\frac{\partial}{\partial t}(\rho u^i u^j) = -u^i u^j \frac{\partial}{\partial x^k}(\rho u^k) - u^k \frac{\partial}{\partial x^k} (u^i u^j)\rho- \nonumber \\
&& - u^j \frac{\partial}{\partial x^i} \left( \rho \bar \theta \right) - u^i \frac{\partial}{\partial x^j} \left( \rho \bar \theta \right)= \nonumber \\
&&= -\frac{\partial}{\partial x^k} (\rho u^i u^j u^k) - u^j \frac{\partial}{\partial x^i} \left( \rho \bar \theta \right) - u^i \frac{\partial}{\partial x^j} \left( \rho \bar \theta \right). \label{d/dtnu^iu^j}
\end{eqnarray}
Thus the temporal evolution,
\begin{eqnarray*}
&&\frac{\partial}{\partial t} \left( \rho \bar \theta \delta^{ij} + \rho u^i u^j \right) = -\frac{\partial}{\partial x^k} \left( \rho \bar \theta u^k \right) \delta^{ij} -\frac{2}{D} \rho \bar \theta   \frac{\partial u^k}{\partial x^k}\delta^{ij} -\nonumber \\
&& - \frac{\partial}{\partial x^k}(\rho u^i u^j u^k) - u^j \frac{\partial}{\partial x^i} \left( \rho \bar \theta   \right) - u^i \frac{\partial}{\partial x^j} \left( \rho \bar \theta   \right),
\end{eqnarray*}
is to be fed into Eq.(\ref{master_eq_momentum}) to allow for the substitution of the position and time cross derivative term:
\begin{eqnarray}
&&\frac{\partial}{\partial t} \left( \rho \bar \theta \delta^{ij} + \rho u^i u^j \right) = -\frac{\partial}{\partial x^k} \left[ \rho u^k \left( \bar \theta  \delta^{ij} + u^i u^j\right) \right]- \nonumber \\
&& -\frac{2}{D} \rho \bar \theta \frac{\partial u^k}{\partial x^k} \delta^{ij} -u^j \frac{\partial}{\partial x^i}\left( \rho \bar \theta \right) - u^i\frac{\partial}{\partial x^j} \left( \rho \bar \theta \right). \label{momentum_crossing_derivative_term}
\end{eqnarray}
Finally we are able to write the addition of two terms proportional
to $\tau$ in Eq. (\ref{eq_momentum_with_crossing_derivative}) in
terms of the derivative of the viscosity stress tensor
(Eq.(\ref{sigma_definition})) by means of
Eq.(\ref{momentum_crossing_derivative_term}).
\begin{eqnarray}
&&\frac{\partial}{\partial x^j} \frac{\partial}{\partial x^k} \left[ \rho(u^k \delta^{ij}+ u^j \delta^{ik} + u^i\delta^{jk})\bar \theta + \rho u^i u^j u^k\right]+ \nonumber\\
&&+\frac{\partial}{\partial x^j} \frac{\partial}{\partial t} \left[ \rho \left( \bar \theta   \delta^{ij} + u^i u^j\right) \right]=  \nonumber \\
&&= \frac{\partial}{\partial x^j} \frac{\partial}{\partial x^i} \left( \rho u^j \bar \theta   \right) + \frac{\partial}{\partial x^j} \frac{\partial}{\partial x^j} \left( \rho \bar \theta u^i   \right) - \frac{\partial}{\partial x^i} \left(\frac{2}{D} \rho \bar \theta   \frac{\partial u^k}{\partial x^k} \right)- \nonumber \\
&&- \frac{\partial}{\partial x^j} \left[ u^j \frac{\partial}{\partial x^i}\left( \rho \bar \theta  \right) \right] - \frac{\partial}{\partial x^j} \left[ u^i \frac{\partial}{\partial x^j} \left( \rho \bar \theta \right) \right]= \nonumber \\
&&= \frac{\partial}{\partial x^j} \left( \rho \bar \theta \frac{\partial u^j}{\partial x^i}\right) + \frac{\partial}{\partial x^j} \left( \rho \bar \theta \frac{\partial u^i}{\partial x^j} \right) - \frac{\partial}{\partial x^i} \left( \frac{2}{D} \rho \bar \theta   \frac{\partial u^k}{\partial x^k} \right)= \nonumber \\
&&= \frac{\partial}{\partial x^j} \left[ \rho \bar \theta \left( \frac{\partial u^j}{\partial x^i}+ \frac{\partial u^i}{\partial x^j} -\frac{2}{D} \frac{\partial u^k}{\partial x^k} \delta^{ij} \right) \right].
\end{eqnarray}
Substituting this result in Eq.(\ref{eq_momentum_with_crossing_derivative}), we obtain the momentum conservation equation of Eq.(\ref{navierstokeseq}).

\subsection{The energy conservation equation}
Some identities are obtained here, similarly to the momentum
equation case. From Eqs.(\ref{eq_momentum+continuity}) and
(\ref{d/dtnthetaDg/g}), it follows that,
\begin{eqnarray}\label{d/dt(D+2)nthetag/gui}
&&\frac{\partial}{\partial t} \left[ (D+2) \rho \bar \theta u^j \right] = (D+2) u^j \frac{\partial}{\partial t} \left( \rho\bar \theta \right) + (D+2) \rho\bar \theta \frac{\partial u^j}{\partial t}=  \nonumber\\
&&= -(D+2) \Bigg[ \frac{\partial}{\partial x^k} \left( \rho\bar \theta u^j u^k \right) + \frac{2}{D} \rho  \bar \theta   \frac{\partial u^k}{\partial x^k} u^j  + \bar \theta   \frac{\partial}{\partial x^j} \left( \rho\bar \theta \right) \Bigg]. \nonumber\\
\end{eqnarray}
Using the continuity equation (Eq.(\ref{continuity})) and
Eq.(\ref{eq_momentum+continuity}), one obtains that,
\begin{eqnarray}\label{d/dtnu^ju^2}
&&\frac{\partial}{\partial t}(\rho u^j \boldsymbol{u}^2) = \frac{\partial \rho}{\partial t}u^j\boldsymbol{u}^2 +\rho\frac{\partial u^j}{\partial t} \boldsymbol{u}^2 + 2\rho u^j u^k \frac{\partial u^k}{\partial t}=\nonumber \\
&&=-\frac{\partial}{\partial x^k}\left( \rho u^k u^j \boldsymbol{u}^2 \right) - \frac{\partial}{\partial x^j} \left( \rho \bar \theta \right) \boldsymbol{u}^2 -2u^j u^k \frac{\partial}{\partial x^k}\left( \rho \bar \theta \right). \nonumber \\
\end{eqnarray}
Eqs.(\ref{d/dt(D+2)nthetag/gui}) and (\ref{d/dtnu^ju^2}) allows to
write the time dependent term of the position-time cross derivative
in Eq. (\ref{macro-e2}) as,
\begin{eqnarray}\label{last_term_in_eq_of_energy}
&&\frac{\partial}{\partial t} \left[ \rho \boldsymbol{u}^2 u^j + \rho \bar \theta (D+2) u^j\right] =\nonumber \\
&&-\frac{\partial}{\partial x^k}\left[\rho u^k u^j \boldsymbol{u}^2 + (D+2)\rho \bar \theta u^j u^k  \right] -\frac{2}{D}(D+2) \rho\bar \theta  \frac{\partial u^k}{\partial x^k}u^j- \nonumber\\
&&-\left[(D+2)\bar \theta   + \boldsymbol{u}^2\right] \frac{\partial}{\partial x^j}\left(\rho\bar \theta \right)-2u^j u^k \frac{\partial}{\partial x^k}\left( \rho \bar \theta \right).
\end{eqnarray}
Next we deal with the terms proportional to $\tau$ in {\it the left
side} of Eq.(\ref{macro-e2}) and apply
Eq.(\ref{last_term_in_eq_of_energy}) to it. For this define,
\begin{eqnarray}\label{Rdefinition}
&&\tau\left[ \cdots \right] \equiv \frac{1}{2}\frac{\partial}{\partial x^j} \frac{\partial}{\partial x^k} \left[ \rho \boldsymbol{u}^2 u^j u^k+ \rho \bar \theta  (D+4) u^j u^k + \right .\nonumber \\
&&\left . +\rho \bar \theta  \boldsymbol{u}^2 \delta^{jk}  + \rho \bar \theta^2 (D+2)\delta^{jk} \right]+ \nonumber \\
&&+\frac{1}{2}\frac{\partial}{\partial x^j} \frac{\partial}{\partial t}\left[ \rho u^j \boldsymbol{u}^2 + \rho \bar \theta  (D+2) u^j \right],
\end{eqnarray}
and substitute (\ref{last_term_in_eq_of_energy}) into the above
equation plus the cancelling of some terms, giving the final useful
expression for the $\tau$ dependent terms of {\it the left side} of
Eq.(\ref{macro-e2}):
\begin{eqnarray}\label{tau-aux}
&&\tau\left[ \cdots \right]= \frac{D+2}{2} \frac{\partial}{\partial x^j} \left( \rho{\bar \theta}\frac{\partial}{\partial x^j} {\bar \theta} \right) + \nonumber \\
&& + \frac{\partial}{\partial x^j}\left[ \rho\bar \theta u^k \left( \frac{\partial u^k}{\partial x^j} + \frac{\partial u^j}{\partial x^k} -\frac{2}{D} \frac{\partial u^l}{\partial x^l}\delta^{jk} \right) \right].
\end{eqnarray}
The above expression is introduced into Eq.(\ref{macro-e2}), added
to the definitions of the dynamic viscosity
(Eq.(\ref{eta_definition})), the viscosity stress tensor
(Eq.(\ref{sigma_definition})) and the thermal conductivity
(Eq.(\ref{kappa_definition})).


\section{Useful Relations}\label{usefulrelations}

\begin{equation}
\sum_\alpha \xi^i_\alpha \xi^j_\alpha \xi_\alpha^k \xi^k_\alpha w_\alpha =(D+2)\delta^{ij}
\label{ur1}
\end{equation}
\begin{equation}
\sum_\alpha \xi^i_\alpha \xi^j_\alpha (\boldsymbol{\xi_\alpha \cdot u})^2 w_\alpha = \boldsymbol{u}^2 \delta^{ij} + 2u^iu^j
\label{ur2}
\end{equation}
\begin{equation}
\sum_\alpha \xi^i_\alpha \xi_\alpha^j (\boldsymbol{\xi_\alpha \cdot u})^4 w_\alpha = 3 \boldsymbol{u}^4 \delta^{ij} + 12 \boldsymbol{u}^2 u^i u^j
\label{ur3}
\end{equation}
\begin{equation}
\sum_\alpha \xi_\alpha^i \xi_\alpha^j (\boldsymbol{\xi_\alpha \cdot u})^2 \boldsymbol{\xi_\alpha}^2 w_\alpha = u^i u^j (2D+8)+ \delta^{ij}(D+4) \boldsymbol{u}^2
\label{ur4}
\end{equation}
\begin{equation}
\sum_\alpha \xi_\alpha^i \xi_\alpha^j \boldsymbol{\xi_\alpha}^4 w_\alpha = \delta^{ij} (D^2+ 6D+8)
\label{ur5}
\end{equation}
\begin{equation}
\sum_\alpha \xi_\alpha^i \xi_\alpha^j \xi_\alpha^k (\boldsymbol{\xi_\alpha \cdot u })w_\alpha = u^k \delta^{ij} + u^j \delta^{ik}+ u^i \delta^{jk}
\label{ur6}
\end{equation}
\begin{equation}
\sum_\alpha \xi_\alpha^i \xi_\alpha^j \xi^k_\alpha (\boldsymbol{\xi_\alpha \cdot u  })^3 w_\alpha = 3(u^k\delta^{ij}+ u^j \delta^{ik}+u^i\delta^{jk}) \boldsymbol{u}^2 + 6u^i u^j u^k
\label{ur7}
\end{equation}
\begin{equation}
\sum_\alpha \xi_\alpha^i \xi_\alpha^j \xi_\alpha^k (\boldsymbol{\xi_\alpha \cdot u}) \boldsymbol{\xi_\alpha}^2 w_\alpha = (D+4) (u^k\delta^{ij}+ u^j \delta^{ik}+u^i\delta^{jk})
\label{ur8}
\end{equation}
\begin{equation}
\sum_\alpha \boldsymbol{\xi_\alpha}^2 w_\alpha =D
\label{ur9}
\end{equation}
\begin{equation}
\sum_\alpha \boldsymbol{\xi_\alpha}^2(\boldsymbol{\xi_\alpha \cdot u})^2 w_\alpha =(D+2)\boldsymbol{u}^2
\label{ur10}
\end{equation}
\begin{equation}
\sum_\alpha \boldsymbol{\xi_\alpha}^2 (\boldsymbol{\xi_\alpha \cdot u})^4 w_\alpha = 3(D+4)\boldsymbol{u}^4
\label{ur11}
\end{equation}
\begin{equation}
\sum_\alpha \boldsymbol{\xi_\alpha}^4 w_\alpha = D(D+2)
\label{ur12}
\end{equation}
\begin{equation}
\sum_\alpha \boldsymbol{\xi_\alpha}^4 (\boldsymbol{\xi_\alpha \cdot u})^2 w_\alpha = (D+2)(D+4)\boldsymbol{u}^2
\label{ur13}
\end{equation}
\begin{equation}
\sum_\alpha \boldsymbol{\xi_\alpha}^6 w_\alpha = D^3 + 6D^2 +8D
\label{ur14}
\end{equation}
\begin{equation}
\sum_\alpha \boldsymbol{\xi_\alpha}^2 \xi_\alpha^i (\boldsymbol{\xi_\alpha \cdot u}) w_\alpha = (D+2) u^j
\label{ur15}
\end{equation}
\begin{equation}
\sum_\alpha \boldsymbol{\xi_\alpha}^2 \xi_\alpha^j (\boldsymbol{\xi_\alpha \cdot u})^3 w_\alpha = 3(D+4) u^j \boldsymbol{u}^2
\label{ur16}
\end{equation}
\begin{equation}
\sum_\alpha \boldsymbol{\xi_\alpha}^4 \xi_\alpha^j (\boldsymbol{\xi_\alpha \cdot u})w_\alpha = (D+4)(D+2) u^j
\label{ur17}
\end{equation}
\begin{equation}
\sum_\alpha \boldsymbol{\xi_\alpha}^2 \xi_\alpha^j \xi_\alpha^k (\boldsymbol{\xi_\alpha \cdot u})^4 w_\alpha = 3(D+6)(\boldsymbol{u}^4 \delta^{jk} + 4u^j u^k \boldsymbol{u}^2)
\label{ur18}
\end{equation}
\begin{equation}
\sum_\alpha \boldsymbol{\xi_\alpha}^4 \xi_\alpha^j \xi_\alpha^k (\boldsymbol{\xi_\alpha \cdot u})^2 w_\alpha = (D+4)(D+6)(\boldsymbol{u}^2 \delta^{jk} + 2u^j u^k)
\label{ur19}
\end{equation}
\begin{equation}
\sum_\alpha \boldsymbol{\xi_\alpha}^6 \xi_\alpha^j \xi_\alpha^k w_\alpha = (D+2)(D+4)(D+6) \delta^{jk}
\label{20}
\end{equation}

\end{document}